\documentclass[fleqn,usenatbib]{mnras}

\usepackage[T1]{fontenc}

\DeclareRobustCommand{\VAN}[3]{#2}
\let\VANthebibliography\thebibliography
\def\thebibliography{\DeclareRobustCommand{\VAN}[3]{##3}\VANthebibliography}

\usepackage{graphicx}	% Including figure files

\usepackage{natbib}
\bibpunct{(}{)}{;}{a}{}{,}
\usepackage{txfonts}
\usepackage{xcolor}
\usepackage{multirow}

\usepackage{epsfig}
\usepackage{longtable}
\usepackage{lscape}
\usepackage{refcount}
\def\inte{{\em INTEGRAL}}
\def\xmm{{\em XMM-Newton}}

\def\rxte{{\em RXTE}}
\def\swift{{\em Swift}}

\def\suzaku{{\em Suzaku}}
\def\nustar{{\em NuSTAR}}

\def \srca {IGR~J16418$-$4532}
\def \srcb {IGR~J16479$-$4514}
\def \srcc {IGR~J16493$-$4348}
\def \mysrc {2S~0114+650} 

\def \ATel {The Astronomer's Telegram}

 \definecolor{arancio}{rgb}{1,0.5,0}
 \definecolor{viola}{rgb}{0.7,0,1}
 \definecolor{verde}{rgb}{0.2,0.7,0.7}
\definecolor{cobalt}{rgb}{0.0, 0.28, 0.67}
\definecolor{airforceblue}{rgb}{0.36, 0.54, 0.66}
\definecolor{ballblue}{rgb}{0.13, 0.67, 0.8}
\definecolor{battleshipgrey}{rgb}{0.52, 0.52, 0.51}
\definecolor{darkgreen}{rgb}{0.0, 0.2, 0.13}

\title[Superorbital modulations in SG X-ray binaries]{{\em Swift}/XRT observations of superorbital modulations in wind-fed supergiant X-ray binaries }

\author[Romano et al.]{
P.\ Romano,$^{1}$\thanks{E-mail: patrizia.romano@inaf.it}
E.\ Bozzo,$^{2,}$$^{3}$
N.\ Islam, $^{4,}$$^{5}$
R.H.D. Corbet, $^{6,}$$^{5,}$$^{7}$
\\
$^{1}$INAF, Osservatorio Astronomico di Brera, Via E.\ Bianchi 46, I-23807, Merate, Italy \\ 
$^{2}$Department of Astronomy, University of Geneva, Chemin d'Ecogia 16, CH-1290 Versoix, Switzerland\\
$^{3}$INAF-OAR, Via Frascati, 33, 00078 Monte Porzio Catone, Rome, Italy\\ 
$^{4}$Center for Space Science and Technology, University of Maryland, Baltimore County, 1000 Hilltop Circle, Baltimore, MD 21250, USA \\ 
$^{5}$X-ray Astrophysics Laboratory, NASA Goddard Space Flight Center, Greenbelt, MD 20771, USA\\ 
$^{6}$CRESST and CSST, University of Maryland, Baltimore County, 1000 Hilltop Circle, Baltimore, MD 21250, USA\\ 
$^{7}$Maryland Institute College of Art, 1300 W Mt Royal Ave, Baltimore, MD 21217, USA\\ 
}

\date{}

\begin{document}
\label{firstpage}
\pagerange{\pageref{firstpage}--\pageref{lastpage}}
\maketitle

\begin{abstract}
We present the first {\it Swift}/XRT long-term monitoring of 2S~0114+650, a  wind-fed supergiant X-ray binary for which both orbital and superorbital periods are known 
(P$_{\rm orb}\sim$11.6\,d and P$_{\rm sup}\sim$30.8\,d). 
Our campaign, summing up to $\sim 79$\,ks,  
is the most intense and complete sampling of the X-ray light curve of this source with a sensitive pointed X-ray instrument, and covers 
17 orbital, and 6 superorbital cycles.  
The combination of flexibility, sensitivity, and soft X-ray coverage of XRT allowed us to confirm previously reported 
spectral changes along the orbital cycle of the source and unveil the variability in its spectral parameters as a 
function of the superorbital phase. 
For completeness, we also report on a similar analysis carried out by exploiting XRT archival data on three additional wind-fed supergiant 
X-ray binaries IGR~J16418$-$4532, IGR~J16479$-$4514, and IGR~J16493$-$4348. For these sources, the archival data provided coverage along 
several superorbital cycles but our analysis could not reveal any significant spectral variability.  
\end{abstract}


   %%%%%%%%%%%%%%%%%%%%%%%%%%%%%%%%%%%%%%%%%%%%%%%%%%%%%
   \section{Introduction}\label{supero:intro}
   %%%%%%%%%%%%%%%%%%%%%%%%%%%%%%%%%%%%%%%%%%%%%%%%%%%%%

Superorbital modulations have been detected in the past decade from several different classes of X-ray
binaries, containing both neutron star (NS) and black hole accretors. These modulations are revealed by
exploiting long-term (typically months to years) monitoring observations of relatively bright systems in
the X-rays and usually appear as significant peaks in the system power density spectra with
periodicities that can be as long as several to tens of times those associated with the binary orbital
period \citep[see, e.g.,][for a historical review]{Sood2007}. In those X-ray binaries where the
accretion takes place via an accretion disk around the compact object, the superorbital modulation is
commonly ascribed to the precession of a (sometimes) warped disk. The precession model has been
extensively applied to systems like Her\,X-1, where spectroscopic investigations in the X-ray domain
were able to constrain in fairly good detail the geometry of the X-ray emission along an entire
superorbital cycle \citep[see, e.g.,][and references therein]{Brumback2021:herx1last}. The precessing
accretion disk model has also been able to convincingly explain the superorbital modulations observed
from LMC\,X-4 \citep{Ambrosi2022} and SMC\,X-1 \citep{Brumback2020,Pradhan2020}.

It proved considerably more complicated to explain the superorbital modulations revealed from wind-fed
X-ray binaries, where the accretion onto the compact object takes place directly from the strong wind of
a massive companion (without the presence of a disk). Wind-accreting X-ray binaries are generally
fainter than disk-fed systems and thus revealing superorbital modulations requires years of monitoring
observations carried out with large field-of-view (FOV) instruments \citep[e.g., those on-board \rxte,\
\inte,\ and \swift; see][for a recent review]{CorbetKrimm2013:superorbital}. In this class of sources,
alternative models for the superorbital modulations have been proposed. \citet{Koenigsberger2003} and
\citet{Moreno2005} showed that oscillations induced in non-synchronously rotating stars in binary
systems could lead to changes in the mass loss rates (and thus on the accretion-driven X-ray luminosity)
on periods longer than the binary orbital period. Alternatively, the presence of a third body in a more
distant orbit but gravitationally bound to the binary could also produce a periodic long-term variation
of the X-ray luminosity \citep[see, e.g.][and references therein]{Farrell2008}. More recently,
\citet{Bozzo2017:superorbital} proposed that the superorbital modulation could be associated with the
periodic interaction between the accreting compact object and the so-called corotating interaction
regions (CIRs) around the massive companion. These structures are known to extend for several stellar
radii around OB supergiants, being characterized by a substantial over-density compared to the
surrounding stellar wind and a possibly asynchronous velocity with respect to the massive star rotation.
Due to their physical properties, the interaction between the NS and the CIRs is expected to lead not
only to an X-ray intensity variation but also to changes (by a factor of a few) of the absorption column
density local to the source \citep[see also][and references therein]{Lobel2008}.
 
Currently, only six wind-fed binaries are known to show superorbital modulations, IGR~J16479$-$4514,
IGR~J16418$-$4532, 2S~0114+650, 4U~1538$-$522, 4U~1909+07, and IGR~J16493$-$4348. All of them displayed
a puzzling virtually identical ratio between the orbital and superorbital periodicity of a factor of
2.7--4  \citep{CorbetKrimm2013:superorbital,Coley2019:16493superorbital,Corbet2021:superorbital,Islam2023_superorbital}. 
Attempts have been made in the literature to interpret their superorbital modulations in the context of
the three models described above. \citet{CorbetKrimm2013:superorbital} argued that the applicability of
the stellar oscillations and the third body model are rather difficult. The former scenario requires
circular orbits \citep[see also][]{Moreno2011} and it is hardly compatible with the high coherency of
the superorbital modulations. The second scenario could be compatible with the high coherency but the
measured virtually constant ratio between the orbital and superorbital periodicity of a factor of
$\sim$2.7--4 across different sources would require unrealistic constraints on their formation
conditions (especially in view of the expectation that these should eventually be hierarchical systems
hosting a distant third body). \citet{Corbet2021:superorbital} and \citet{Islam2023_superorbital}
discussed how the CIR model could provide a more likely explanation for most sources, with the virtually
constant ratio between the system orbital and superorbital modulations associated with a
pseudo-synchronization of the CIR rotational period with the NS orbit.

Investigations of the origin of the superorbital modulations in wind-fed X-ray binaries in the X-ray
domain have so far relied on two kinds of observational campaigns: (i) long term monitoring provided
(mostly) at hard X-rays by large FOV instruments, and (ii) relatively short (few tens of ks) pointed
observations per object during a specific superorbital phase. Monitoring observations proved so far
crucial to discover the periodicities, although they generally provide limited information on possible
spectral variations along the superorbital cycle due to the relatively low sensitivity and
signal-to-noise ratio (S/N) over short time scales \citep[see, e.g.,][and references
therein]{CorbetKrimm2013:superorbital,Corbet2017}. Pointed observations carried out also with multiple
facilities at the same time are characterized by a much higher sensitivity and allow in-depth analysis
of the intensity and spectral energy distribution of the source. However, they are affected by the
limitation of providing only a snapshot of the binary system during a relatively short time interval and
it is thus difficult to unambiguously link any measured variability of the source intensity or of any
identified spectral feature with the superorbital phase. As discussed by \citet{Bozzo2017:superorbital},
wind-fed binaries are commonly characterized by a prominent intensity and spectral variability (on time
scales of hundreds of seconds to hours) due to the clumpiness of the stellar wind \citep[see also][for
recent reviews]{Nunez2017,Kretschmar2019} from the companion star. A proper study of any source
variability with the superorbital phase thus requires long-term observations carried out across many
cycles with high sensitivity, so that they can be folded over the superorbital period and ensure the
short-term variability due to the stellar wind clumps is averaged out.

In the past, the process of averaging X-ray observations has been exploited to unveil the presence of
long-lived structures surrounding the accreting compact objects in wind-fed X-ray binaries along their
orbital rather then their superorbital cycles. For very bright X-ray sources as Vela X-1, GX 301-2, and
4U 1538-52, the data collected by the Monitor of All Sky X-ray Image
\citep[][MAXI]{MAXI:Matsuoka2009PASJmn} proved particularly interesting due to their extended energy
coverage over many orbital revolutions since the beginning of the experiment scientific operations back
in 2009 \citep[][]{Doroshenko2013:velax1,Islam2014,Rodes-Roca2015:1538}. However, the vast majority of
the wind-fed X-ray binaries are either too faint or located into crowded sky regions to be efficiently
accessed by the MAXI capabilities. For all these cases, our group has extensively demonstrated that
observations with the narrow-field instrument, the X--ray Telescope \citep[XRT,][]{Burrows2005:XRT}
on-board the Neil Gehrels \swift\ Observatory \citep[][]{Gehrels2004}, are very well suited
\citep{Ferrigno2022:SGXRB}.

In this paper, we present for the first time the exploitation of XRT data to probe spectral variability
not only on the orbital but also along the superorbital cycles of wind-fed X-ray binaries. In
particular, we report on the outcomes of a new long-term XRT observational campaign focused on
2S~0114+650. The campaign was specifically designed to unveil spectral and intensity variability of this
wind-fed binary along its superorbital cycle. Thanks to the regular monitoring and the enhanced
sensitivity of \swift/XRT, we can now simultaneously average data over multiple orbital and superorbital
cycles, a crucial step in mitigating short-term X-ray intensity variations and exploring long-term
spectral changes. We also report the results obtained by applying the same techniques as those exploited
in the case of 2S~0114+650 on archival data collected from three further wind-fed binaries,
IGR\,J16418-4532, IGR\,J16479-4514, and IGR \,J16493-4348 (see Table~\ref{supero:tab:sample}). For these
objects the past observational campaigns, albeit not optimized for the search of variability associated
to the superorbital period, contained sufficient data to apply our method. A brief summary of the
current knowledge on all sources we considered is given in Sect.~\ref{supero:sources}. In
Sect.~\ref{supero:dataredu}, we provide a description of the \swift/XRT data (with log tables included
in Appendix~\ref{supero:appendix1}) and the data-reduction technique. An overview of the results we
obtained and their discussion are provided in Sect.~\ref{supero:discussion}.

            %%%%%%%%%%%%%%%%%%%%%%%%%%%%%%%%%%%%%%%%%%%%%%%%%% TABLE 1
  \begin{table}
\tabcolsep 2pt
 \begin{center}
 \caption{Properties of the sources studied in this paper. \label{supero:tab:sample} }
  \begin{tabular}{rcccccc}
\noalign{\smallskip}
 \hline
 \noalign{\smallskip}
Source                                                  & P$_{\rm orb}$                         &   P$_{\rm sup}$   &                                                 &  \multicolumn{2}{c}{References} \\
                                                          &                          &                  &                  $T_{0}$$^a$                                      & P$_{\rm orb}$        &   P$_{\rm sup}$ \\
                                                             & (d)                                 &    (d)                   &                 (MJD)                                        & \\
\noalign{\smallskip}
 \hline
 \noalign{\smallskip}
2S~0114+650                    &  11.5983$\pm$0.0006$^b$ &   30.76$\pm$0.03    &   53488                       &  \getrefnumber{tabGrundstrom2007}  &  \getrefnumber{tabCorbetKrimm2013:superorbital}   \\ 
IGR~J16418$-$4532         &  3.73886$\pm$0.00003  &  14.730$\pm$0.006 &   55994.6                & \getrefnumber{tabLevine2011}                        & \getrefnumber{tabCorbetKrimm2013:superorbital} \\
IGR~J16479$-$4514         &  3.3193$\pm$0.0005       &  11.880$\pm$0.002 &   55996                   &  \getrefnumber{tabRomano2009:sfxts_paperV}              & \getrefnumber{tabCorbetKrimm2013:superorbital} \\
IGR~J16493$-$4348           &  6.782$\pm$0.005           &  20.07$\pm$0.01     &   54265.1                &  \getrefnumber{tabCusumano2010:16493-4348_period} & \getrefnumber{tabCorbetKrimm2013:superorbital},\getrefnumber{tabColey2015:eclipsing_binaries} \\
\noalign{\smallskip}
\hline
\noalign{\smallskip}
\end{tabular}            

References:
\newcounter{ctr_tabrefs}
\setrefcountdefault{-99}
 \refstepcounter{ctr_tabrefs}\label{tabGrundstrom2007}(\getrefnumber{tabGrundstrom2007})  \citet{Grundstrom2007}; 
\refstepcounter{ctr_tabrefs}\label{tabCorbetKrimm2013:superorbital}(\getrefnumber{tabCorbetKrimm2013:superorbital})  \citet{CorbetKrimm2013:superorbital};  
\refstepcounter{ctr_tabrefs}\label{tabLevine2011}(\getrefnumber{tabLevine2011})  \citet{Levine2011};
    \refstepcounter{ctr_tabrefs}\label{tabRomano2009:sfxts_paperV}(\getrefnumber{tabRomano2009:sfxts_paperV})  \citet{Romano2009:sfxts_paperV};
  \refstepcounter{ctr_tabrefs}\label{tabCusumano2010:16493-4348_period}(\getrefnumber{tabCusumano2010:16493-4348_period})  \citet{Cusumano2010:16493-4348_period};  
 \refstepcounter{ctr_tabrefs}\label{tabColey2015:eclipsing_binaries}(\getrefnumber{tabColey2015:eclipsing_binaries})  \citet{Coley2015:eclipsing_binaries}.

Notes: $^a$ Phase zero for the superorbital period is the minimum of the folded light curve for 2S~0114+650 \citet[][]{Farrell2008},  
the maximum for the remainder of the sample.
$^b$ Phase zero for the orbital period is the time of periastron passage from \citet[][]{Grundstrom2007} at MJD 51824.8.
\end{center}
\end{table}

    %%%%%%%%%%%%%%%%%%%%%%%%%%%%%%%%%%%%%%%%%%%%%%%%%%%%%
    \section{The source sample} \label{supero:sources}
   %%%%%%%%%%%%%%%%%%%%%%%%%%%%%%%%%%%%%%%%%%%%%%%%%%%%%

In the sub-sections below, we briefly describe the current knowledge on all sources considered in this paper. 
We focus more on 2S~0114+650, which is the target of our recent observational campaign with \swift/XRT. 
The other three sources (IGR~J16418-4532, IGR~J16479-4514, and IGR~J16493-4348) are briefly mentioned 
for completeness, as we apply to their archival data (already analyzed in previous work) the same analysis 
techniques used in the case of 2S~0114+650.

    %%%%%%%%%%%%%%%%%%%%%%%%%%%%%%%%%%%%%%%%%%%%%%%%%%%%%
    \subsection{2S~0114+650} \label{supero:2s}
    %%%%%%%%%%%%%%%%%%%%%%%%%%%%%%%%%%%%%%%%%%%%%%%%%%%%%

2S~0114+650 was discovered in 1977 and readily classified as a supergiant high-mass X-ray 
binary, with the companion star being identified as a B1Ia supergiant (LS I+65 010) at a 
distance of $\sim$7.2 kpc \citep[SgXB;][]{Dower1977,Reig1996}. 
The nature of the accretor was confirmed as a NS with the discovery of a 2.73\,h spin 
period \citep[see, e.g.][and references therein]{Hall2000}. Subsequent measurements of 
the source spin period also led to the discovery and monitoring of spin-up and spin-down 
phases \citep{Falanga2015}. The measured orbital period of the system is $\sim$11.6~d, 
while the estimated eccentricity is $\sim$0.18 \citep{Koenigsberger2006}. 

The profile of the X-ray variability along the binary orbit shows 
the typical modulation expected for a wind-fed system \citep{Grundstrom2007} but it 
is also characterized by a peculiar stable dip at the inferior conjunction which is most 
likely associated with a localized increase in the absorption column density rather than an 
X-ray eclipse \citep[see][and references therein]{Pradhan2015}. 

S~0114+650 also exhibits a 30.76\,d superorbital modulation of its X-ray emission, which origin has
long been investigated but still remains unclear \citep{Farrell2006,CorbetKrimm2013:superorbital}.
\citet{Hu2017} performed a detailed study of the long-term evolution of the spin, orbital, and
superorbital modulations of the source by using \swift/BAT and \rxte/ASM data. These authors found that
the NS spin period undergoes both episodes of intense and prolonged spin-up/spin-down, and episodes of
more irregular and moderate variations. The orbital and superorbital periodicities were found to be
stable over time, although the intensities of these modulations can vary substantially and generally the
superorbital one decreases during times when the NS spin period variations are more erratic.
\citet{Hu2017} suggested that a temporary accretion disk could form around the NS in 2S~0114+650, such
that the source switches periodically from wind to disk accretion. During the disk accreting periods,
the spin-up/spin-down is stronger and more regular, while during wind accretion it turns into a more
complex behavior, as expected due to the clumpiness of the stellar wind affecting the accretion. The
weakening of the superorbital modulation intensity is more pronounced during the time intervals when the
source is suspected to switch to the wind accretion. The authors thus advanced the hypothesis that the
modulation itself could be somehow closely linked to the presence of an accretion disk.

The broad-band X-ray spectrum of 2S~0114+650 has been measured several times in the past decades using
different facilities. The source spectral energy distribution is closely reminiscent of what is usually
observed from wind-fed systems and well described by a model comprising a substantially absorbed
(3--5$\times$10$^{22}$~cm$^{-2}$) cut-off power law. The cut-off energy is found to be in the range
10--30\,keV, while typical power-law photon indices vary between 1.0--2.3
\citep{Farrell2008,Pradhan2015,Abdallah2023}. To the best of our knowledge, a detailed study of the
spectral variability of the source along its orbital and superorbital cycles have been presented in the
past only by \citet{Farrell2008}. These authors used an absorbed cut-off model to describe the source
X-ray emission measured by the \rxte/PCA and revealed a modest change in both the source absorption
column density and photon index as a function of the orbital phase. In particular, they found that the
spectrum is harder and less absorbed when the source emission is fainter. A similar trend for the photon
index and absorption column density was suggested for the superorbital case as well, although the error
bars associated with the measurements were far too large to draw any firm conclusion. No changes were
measured as a function of both the orbital and superorbital period for the cut-off energy.

    %%%%%%%%%%%%%%%%%%%%%%%%%%%%%%%%%%%%%%%%%%%%%%%%%%%%%
    \subsection{The other sources: IGR~J16418-4532, IGR~J16479-4514, and IGR~J16493-4348} \label{supero:others}
    %%%%%%%%%%%%%%%%%%%%%%%%%%%%%%%%%%%%%%%%%%%%%%%%%%%%%
IGR~J16418-4532 has been classified as an intermediate 
supergiant fast X-ray transient \citep[SFXTs; e.g.\ for recent reviews, see][]{Walter2015:HMXBs_IGR,Nunez2017} 
since flares from this source have been observed several times 
\citep[][]{Tomsick2004:atel224,Sguera2006,Ducci2010,Romano2012:sfxts_16418,Krimm2013:atel5398,Romano2014:sfxts_catI,Romano2023:sfxts_catII}
but the overall dynamic range in the X-ray luminosity remains as of today limited to values somewhat 
smaller than those typical of the bulk of the SFXT population \citep{Romano2015:17544sb}. 
The source displays clear eclipses, with a measured orbital period of 3.73~days. The nature of the 
compact object has been firmly established to be a NS thanks to the discovery of X-ray pulsations 
with a period of $\sim$1210~s \citep{Sidoli2012:16418}. 
A precise classification of the donor star is still missing, given the fact that different authors have 
discussed both the possibility of an O or B supergiant located at about 13~kpc 
\citep{Coleiro2013:IR_IDs,Coley2015:eclipsing_binaries}. IGR~J16418-4532 is also known to typically 
display a large absorption column density, exceeding 10$^{23}$~cm$^{-2}$, and a superorbital 
modulation of its X-ray emission with a period of $\sim$14.7~days. The superorbital modulation 
has been investigated in detail by \citet{Islam2023_superorbital}. These authors showed that the 
modulation changes in intensity over a time scales of a few years and discussed the results of 
targeted \nustar\ observations (complemented by simultaneous \swift/XRT pointings). 

IGR J16479–4514 is an eclipsing SFXT with an orbital
period of 3.32 days \citep[][]{Jain2009:16479_period}. The nature of the
compact object is unknown because no pulsations have been
detected in the system. However, its spectrum is similar to that
of accreting HMXB pulsars, suggesting a NS as the putative
compact object. The optical companion is likely to be an O7
star, translating into a distance of 4.5\,kpc to the binary system 
\citep[][]{Chaty2008,Coley2015:eclipsing_binaries}.
In addition to short low-luminosity X-ray flares lasting a few thousand seconds, orbital
phase-locked X-ray flares are present in the light curve, with an
occasional bright X-ray flare lasting few hours and reaching an
X-ray flux of 10$^{-9}$ erg cm$^{-2}$ s$^{-1}$ 
\citep[][]{Romano2008:sfxts_paperII,Sguera2008,Sidoli2013:16479,Sguera2020}. 
These flares suggest the presence of large-scale structures in the stellar wind of the
supergiant star \citep[][]{Bozzo2009:16479-4514outburst,Sguera2020}.

IGR~J16493-4348 is an eclipsing high mass X-ray binary with a 
B0.5 Ia supergiant companion \citep[][]{Nespoli2008:16493,Pearlman2019:16493orb_pars} at $\sim16.1$\,kpc,  
and an orbital period of $\sim 6.8$\,d. The X-ray eclipses displayed by the source are 
known to last about 0.1\,d \citep[][]{Cusumano2010:16493-4348_period,Corbet2010:atel2599}. 
Given the discovery of X-ray pulsations with a period of 1093\,s \citep[][]{Corbet2013:atel2766,Pearlman2019:16493orb_pars} and 
the reported evidence of a resonant cyclotron absorption feature at $\sim 30$\,keV \citep{Dai2011:line_16493}, 
the compact object hosted in the binary system is believed to be a NS with a surface magnetic field of $3.7\times10^{12}$\,G. 
A superorbital modulation of the X-ray light curve was discovered by \citet[][]{Corbet2010:atel2599} 
and investigated in depth by \citet[][]{Coley2019:16493superorbital}.

   %%%%%%%%%%%%%%%%%%%%%%%%%%%%%%%%%%%%%%%%%%%%%%%%%%%%%%%%%
   \section{Data reduction and analysis} \label{supero:dataredu}
   %%%%%%%%%%%%%%%%%%%%%%%%%%%%%%%%%%%%%%%%%%%%%%%%%%%%%%%%%
In order to probe, as a main goal, possible intensity and spectral variability over the superorbital 
revolution of the NS in 2S~0114+650, 
we planned our campaign (Target ID 15874, PI: P.\ Romano) with {\it Swift}/XRT. 
The monitoring campaign consisted of 1\,ks observations performed 3 times a week starting on 2023-02-10. 
For this work we considered data collected until 
2023-08-24, yielding 74 photon counting (PC) mode observations for a total effective exposure time on the target of 78.8\,ks.
We adopt for the orbital period P$_{\rm orb}$=11.5983$\pm$0.0006\,d       
\citep[][with the phase zero being the time of periastron passage at MJD 51824.8]{Grundstrom2007}, and 
and for the superorbital period P$_{\rm sup}$=30.76$\pm$0.03\,d \citep[][]{CorbetKrimm2013:superorbital}, 
with the phase zero being the time of the minimum of the folded light curve, at MJD 53488 \citep[][]{Farrell2008}. 
Our observations, therefore, cover $\sim 17$ orbital, and $\sim 6$ superorbital cycles.

Table~\ref{supero:tab:swift_xrt_log_s0114} reports  
the log of the {\it Swift}/XRT observations, including the ObsID, 
the observation date (MJD of the middle of the observation), 
the calculated orbital and superorbital phase, the start and end times (UTC), the XRT exposure time, 
and also relevant spectral parameters (see later in this section). 

The XRT data were uniformly processed and analyzed using 
({\sc FTOOLS}\footnote{\href{https://heasarc.gsfc.nasa.gov/ftools/ftools_menu.html}{https://heasarc.gsfc.nasa.gov/ftools/ftools\_menu.html.}} v6.29b),
and matching calibration (CALDB\footnote{\href{https://heasarc.gsfc.nasa.gov/docs/heasarc/caldb/caldb_intro.html}{https://heasarc.gsfc.nasa.gov/docs/heasarc/caldb/caldb\_intro.html.}}) files. 
The spectral analysis was performed with {\sc XSPEC} \citep[][]{Arnaud1996:xspec}, 
by using the C-statistics \citet[][]{Cash1979}   
and by adopting an absorbed power-law model with free absorption and photon index. 
The absorption component was modelled with {\sc tbabs})  
with the default \citet[]{Wilms2000} abundances ({\sc abund wilm}) and 
\citet[][]{Verner1996:xsecs} cross sections ({\sc xsect vern}). 
Errors on the spectral parameters are reported at the 90\,\% confidence level (c.l.). 
The \swift/XRT data were analyzed as follows.

             %%%%%%%%%%%%%%%%%%%%%%%%%%%%%%%%%%%%%%%%%%%%%%%%%% TABLE 2
\begin{table}
 \tabcolsep 4pt
 \begin{center}
\caption{Results of the orbital and superorbital phase-resolved spectral analysis for 2S~0114+650. 
 \label{supero:tab:2S_s_o_phase_specfits} }
\vspace{-0.3cm}
 \begin{tabular}{ ccccr }
 \noalign{\smallskip}
 \hline
 \noalign{\smallskip}
    Phase          & $N_{\rm H}$   & $\Gamma$ & F$_{\rm 0.3-10\,keV}$                            &  Cstat                        \cr
     range                  & $10^{22}$   &                     & $10^{-11}$ &     /d.o.f.      \cr 
                                 &  (cm$^{-2}$)  &                     & (erg\,cm$^{-2}$\,s$^{-1}$) &          \cr 

\noalign{\smallskip} 
\hline   
\noalign{\smallskip}
 Orbital & & & & \\ 
 \hline
 \noalign{\smallskip}
0.00--0.15 & $2.4^{+0.4}_{-0.3}$ & $0.8^{+0.1}_{-0.1}$  & $7.6^{+0.4}_{-0.4}$   &     674.8/694\\
0.15--0.26 & $2.5^{+0.4}_{-0.4}$ & $0.7^{+0.1}_{-0.1}$  & $14.9^{+0.9}_{-0.8}$   &     590.9/675\\
0.26--0.39 & $3.8^{+0.5}_{-0.5}$ & $0.9^{+0.2}_{-0.2}$  & $8.6^{+0.5}_{-0.5}$   &     634.1/664\\
0.39--0.55 & $6.0^{+0.8}_{-0.7}$ & $0.7^{+0.2}_{-0.2}$  & $5.2^{+0.3}_{-0.3}$   &     643.4/664\\
0.55--0.78 & $5.2^{+0.8}_{-0.8}$ & $1.0^{+0.2}_{-0.2}$  & $3.0^{+0.2}_{-0.2}$   &     591.1/616\\
0.78--0.84 & $2.7^{+0.4}_{-0.4}$ & $0.9^{+0.2}_{-0.1}$  & $13.5^{+0.8}_{-0.8}$   &     604.2/658\\
0.84--0.92 & $2.1^{+0.4}_{-0.3}$ & $0.7^{+0.1}_{-0.1}$  & $16.0^{+1.0}_{-0.9}$   &     575.8/638\\
0.92--0.99 & $2.6^{+0.4}_{-0.4}$ & $0.9^{+0.2}_{-0.2}$  & $8.4^{+0.5}_{-0.5}$   &     556.0/613\\
\noalign{\smallskip} 
\hline   
\noalign{\smallskip} 
Superorbital & & & & \\ 
 \hline
 \noalign{\smallskip}
0.000--0.183  & $2.8^{+0.4}_{-0.4}$ & $0.9^{+0.2}_{-0.1}$  & $3.9^{+0.2}_{-0.2}$   &     572.7/647\\
0.183--0.302 & $4.1^{+0.5}_{-0.5}$ & $1.0^{+0.2}_{-0.2}$  & $9.4^{+0.5}_{-0.5}$   &     649.4/656\\
0.302--0.450 & $3.0^{+0.5}_{-0.4}$ & $0.7^{+0.2}_{-0.2}$  & $5.9^{+0.4}_{-0.3}$   &     619.1/667\\
0.450--0.500 & $2.0^{+0.4}_{-0.3}$ & $0.6^{+0.1}_{-0.1}$  & $25.4^{+1.5}_{-1.4}$   &     628.4/692\\
0.500--0.593 & $2.9^{+0.5}_{-0.4}$ & $0.6^{+0.2}_{-0.1}$  & $10.4^{+0.6}_{-0.6}$   &     674.8/673\\
0.593--0.820 & $2.5^{+0.4}_{-0.4}$ & $0.9^{+0.1}_{-0.1}$  & $6.5^{+0.4}_{-0.4}$   &     701.4/673\\
0.820--0.880 & $3.9^{+0.6}_{-0.6}$ & $0.8^{+0.2}_{-0.2}$  & $7.1^{+0.5}_{-0.5}$   &     474.0/610\\
0.880--1.000 & $4.0^{+0.6}_{-0.6}$ & $0.7^{+0.2}_{-0.2}$  & $6.1^{+0.4}_{-0.3}$   &     613.1/645\\
\noalign{\smallskip}
 \hline
 \noalign{\smallskip}
\end{tabular}
 \end{center}
\end{table}

\setcounter{figure}{0}
\begin{figure*} %%%%%%%%%%%%%%%%%%%%%%%%%%%%%%%%%%%%%%%%%%%%%%%%%%%%%%%%% Fig 1

 \hspace{-0.5truecm}
   \includegraphics[width=18.5cm,angle=0]{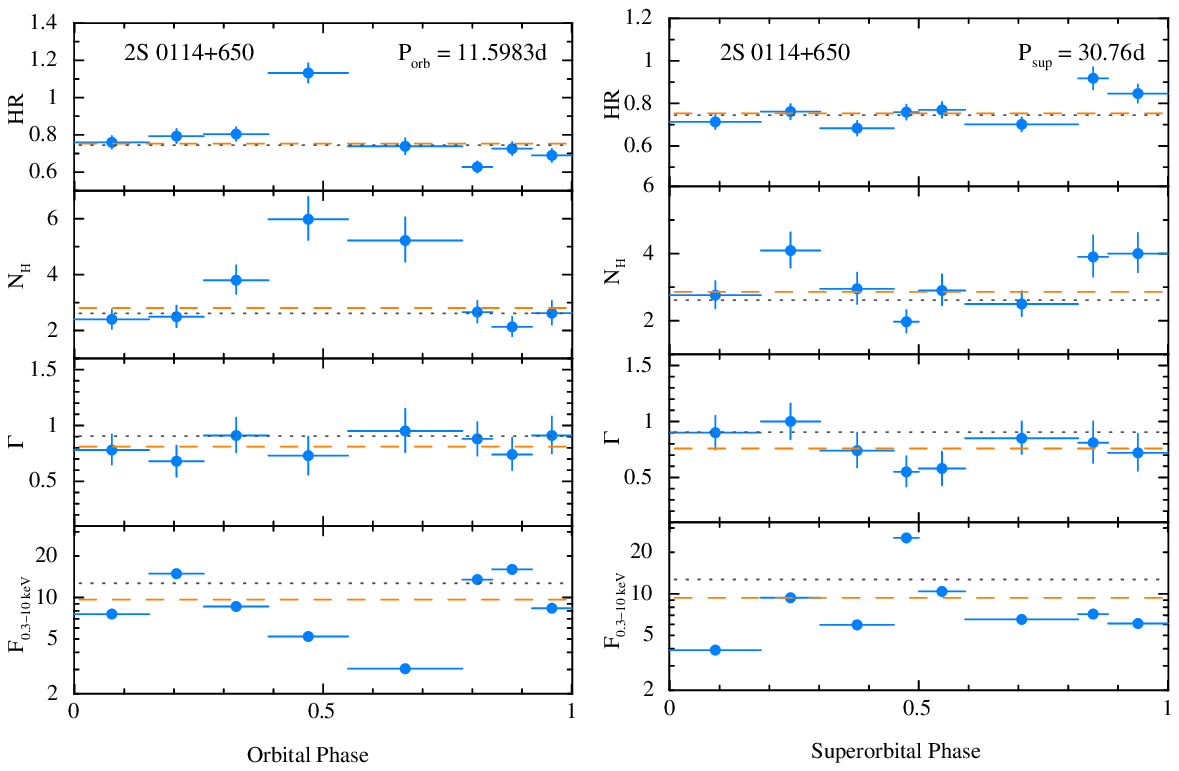}
  
  \vspace{-14.5truecm}
  \caption{{\it Left}: {\it Swift}/XRT hardness ratio of \mysrc, and best-fit parameters as a function of orbital phase 
(P$_{\rm orb}$=11.5983\,d,} $T_{0}$= MJD 51824.8). 
The absorption column density N$_{\rm H}$ is in units of 10$^{22}$ cm$^{-2}$, 
the power-law photon index is $\Gamma$, and the observed 0.3--10 keV flux 
is  in units of 10$^{-11}$ erg cm$^{-2}$ s$^{-1}$). {\it Right}: same as left side but for the superorbital phase 
(P$_{\rm sup}$=30.76\,d, $T_{0}$= MJD 53488, \citealt[][]{CorbetKrimm2013:superorbital}). 
In each panel we also report the mean value of each variable plotted when calculated from the 
 individual observations (dotted line) and from the values obtained in the eight phase bins (orange dashed line). 
The two lines are compatible in all cases within the associated uncertainties.
  \label{supero:fig:2S0114_orbital_phase_fits_campaign}
\end{figure*}%%%%%%%%%%%%%%%%%%%%%%%%%%%%%%%%%%%%%%%%%%%%%%%%%%%%%%%%%

First, for each observation we calculated the orbital phase, and from the count rates in the 
0.3--4 and 4--10\,keV energy bands, we derived the hardness ratio HR = CR(4--10)/CR(0.3--4)  
(also binned by observation).  
We extracted the average spectrum and fitted it with 
an absorbed power law to measure the observed and unabsorbed 0.3--10\,keV flux 
(Cols.\ 8 and 9 of Table~\ref{supero:tab:swift_xrt_log_s0114}).

We then chose eight orbital phase bins that would yield a comparable number of source counts 
(on average $\approx 2070$ cts bin$^{-1}$) and calculated the HR in those bins.  
The top panel of Fig.~\ref{supero:fig:2S0114_orbital_phase_fits_campaign} (left) 
shows the HR in the 8 orbital phase bins, as well as the weighted averages of HRs derived 
both from individual observations (dotted line) and from the 8 bins (dashed line).
We then combined all observations in each of the eight phase bins to create a single spectrum per bin. 
We note that the number of phase bins was chosen as a compromise between obtaining a significant 
number of points along the cycle and maintaining a sufficiently high signal-to-noise ratio spectrum 
to accurately determine both the absorption column density and the power-law slope 
\citep[see also the discussion in our previous paper,][]{Ferrigno2022:SGXRB}. 
Finally, we performed an eight-bin phase-resolved spectroscopic analysis 
in the 0.3--10\,keV energy range adopting 
an absorbed power-law model. 
The results of our orbital-resolved spectral analysis of 2S~0114+650 are summarized in 
Table~\ref{supero:tab:2S_s_o_phase_specfits} and plotted in Fig.~\ref{supero:fig:2S0114_orbital_phase_fits_campaign} (left).

We repeated the same procedure for the creation of superorbital phase bins and the corresponding 
spectroscopic analysis. The results are reported in Table~\ref{supero:tab:2S_s_o_phase_specfits} and 
Fig.~\ref{supero:fig:2S0114_orbital_phase_fits_campaign} (right). 
Finally, in Appendix~\ref{supero:appendix2} we also address the effect of data binning on our result. 

We note that the use of a simple model comprising an absorption component and a power law is 
justified for the source based on previous work (see Sect.~\ref{supero:2s}), due to the limited 
energy band coverage and exposure of the stacked XRT spectra. The addition of a cut-off energy, 
as measured by \rxte\ and \suzaku,\ at energies 10--30\,keV \citep{Farrell2008,Abdallah2023} 
was tested and found not to significantly affect the results obtained here. More complicated models 
including Comptonizing components and reported in a few studies exploiting either the broad-band 
coverage of the instruments on-board \suzaku\ or the deep exposures of the \xmm\ data 
\citep{Pradhan2015,Ferrin2017} were also not usable in the present case, as most parameters would 
simply be unconstrained. As our scope is to unveil the relative variability between observations carried 
out with the same instrument at different orbital and super-orbital phases, the absorbed 
power-law model allows us to unveil if the variability is mostly occurring in the soft X-ray part 
of the spectrum (and thus most likely associated with the absorption column density) or driven 
by changes in the harder part of the XRT energy band (thus more likely associated by a change 
in the overall spectral slope). As a further remark, we note that the procedure we adopted here also 
ensures that the known (modest) spectral energy variability due to the energy-dependent pulsations 
from the source is averaged out in the stacked XRT spectra 
\citep[see][and references therein, and Appendix~\ref{supero:appendix2}]{Pradhan2015}. 
This is because for each orbital and superorbital bin we are merging together many different relatively 
short XRT pointings (typically 1--2~ks at most) over several different orbital and superorbital revolutions 
(see Appendix~\ref{supero:appendix2} for futhr details).

For completeness, we applied the methodology of folding along the superorbital phase 
to three more sources, listed in Table~\ref{supero:tab:sample} along with their relevant properties, 
\srca, \srcb, and \srcc. For these sources, the data available in the {\it Swift} archive were not 
optimized to investigate superorbital spectral variations but rather obtained either as monitoring 
campaigns for the orbital variability or as SFXT discovery/outburst follow-ups \citep[see][]{Romano2023:sfxts_catII}. 
Spectral variations as a function of the orbital phase of these three sources were already 
investigated in previous work \citep[see][and references therein]{Romano2012:sfxts_16418,Sidoli2013:16479,Varun2023:16479_orb_nustar,Pearlman2019:16493orb_pars}. 
In order to optimise the {\it Swift} archival data for superorbital variability studies, 
we selected only observations with homogeneous exposure times, 
so as not to introduce biases in favour of a specific superorbital phase,   
and with an off-axis angle $<15\arcmin$ to limit the known XRT vignetting effects. We then 
applied the same procedures as those adopted for \mysrc, as follows.  

The data we considered for \srca{} are reported in Table~\ref{supero:tab:swift_xrt_log_16418}.   
The majority have mostly been collected either as monitoring campaigns 
\citep[e.g.][]{Romano2012:sfxts_16418} 
or as outburst follow-ups \citep[][and references therein, and in Table~\ref{supero:tab:swift_xrt_log_16418}]{Romano2023:sfxts_catII}.  
Consequently, their exposures vary considerably, 
from $\sim$2--5\,ks obtained during the orbital monitoring in 2011 \citep[][]{Romano2012:sfxts_16418} 
that uniformly cover three orbital periods (1.1 superorbital periods), 
or $\sim$2\,ks triggered observations and their follow-up campaigns 
(typically 1\,ks per observation), to very short serendipitous observations. 
By combining all the data (57 observations for a total exposure of 93\,ks) we obtained   
eight phase-selected spectra with an average of $\approx$ 900 counts each,
the only exception being bin number five, where only one observation 
(00032037022) was available which did not even yield a detection. 

For \srcb, we considered the data published in \citet[][Table~5]{Romano2009:sfxts_paperV} 
and in \citet[][Table~1]{Romano2011:sfxts_paperVI}, that were collected 
to define the long term properties of a sample of SFXTs 
\citep[][]{Sidoli2008:sfxts_paperI,Romano2008:sfxts_paperII,Sidoli2009:sfxts_paperIII,Romano2009:sfxts_paperV,Romano2011:sfxts_paperVI}.
The campaign consisted of 144 observations, each 1\,ks long, obtained twice a week, 
along a baseline of two yearas well as outburst observations and their follow-ups,
for a total of $\sim$160\,ks.  
The observations cover $\sim 54$ superorbital cycles\footnote{We discarded observation 00030296005, 
since it is a single observation dated three months before the remainder of the campaign.}
and yield eight phase-selected spectra with $\approx$ 1650 counts each.

Finally,  for \srcc, we considered the monitoring data we obtained in 2014 
(see Table~\ref{supero:tab:swift_xrt_log_16493}) to study the long term behaviour of this source and 
create the cumulative luminosity distribution, with a cadence of 1\,ks twice a week
for a total of $\sim$53\,ks and 57 observations,  
as reported in \citet[][]{Romano2015:swift10} and \citet[][]{Kretschmar2019}. 
These observations, covering $\sim 11$ superorbital cycles, 
yield eight phase-selected spectra with $\approx 700$ counts each.

Figures~\ref{supero:fig:16418_phase_fits}, \ref{supero:fig:16479_phase_fits}, and \ref{supero:fig:16493_phase_fits} 
show the HR and best-fit parameters as a function of superorbital phase for 
\srca, \srcb, and \srcc, respectively,
while the values of the best fit parameters are reported in Table~\ref{supero:tab:s_o_phase_specfits}.

\begin{figure}%%%%%%%%%%%%%%%%%%%%%%%%%%%%%%%%%%%%%%%%%%%%%%%%%%%%%%%%% FIGURE 2
 \hspace{-0.6truecm}
 \includegraphics[width=9.5cm,angle=0]{figure2.ps}
\vspace{-0.5cm}
  \caption{Same as Fig.~\ref{supero:fig:2S0114_orbital_phase_fits_campaign} (right) but for the case 
of \srca\ (P$_{\rm sup}$=14.730\,d, $T_{0}$= 55994.6 MJD, \citealt[][]{CorbetKrimm2013:superorbital}).}
  \label{supero:fig:16418_phase_fits}
\end{figure}%%%%%%%%%%%%%%%%%%%%%%%%%%%%%%%%%%%%%%%%%%%%%%%%%%%%%%%%%

\begin{figure}%%%%%%%%%%%%%%%%%%%%%%%%%%%%%%%%%%%%%%%%%%%%%%%%%%%%%%%%% FIGURE 3
\hspace{-0.6truecm}
 \includegraphics[width=9.5cm,angle=0]{figure3.ps}
 \vspace{-0.5cm}
 \caption{Same as Fig.~\ref{supero:fig:2S0114_orbital_phase_fits_campaign} (right) but for the case 
of \srcb\ (P$_{\rm sup}$=11.880\,d, $T_{0}$=  MJD 55996, \citealt[][]{CorbetKrimm2013:superorbital}).}
  \label{supero:fig:16479_phase_fits}
\end{figure}%%%%%%%%%%%%%%%%%%%%%%%%%%%%%%%%%%%%%%%%%%%%%%%%%%%%%%%%%

\begin{figure}%%%%%%%%%%%%%%%%%%%%%%%%%%%%%%%%%%%%%%%%%%%%%%%%%%%%%%%%% FIGURE 4
  \hspace{-0.6truecm}
 \includegraphics[width=9.5cm,angle=0]{figure4.ps}
  \vspace{-0.5cm}
 \caption{Same as Fig.~\ref{supero:fig:2S0114_orbital_phase_fits_campaign} (right) but for the case 
of \srcc\ (P$_{\rm sup}$=20.07\,d, $T_{0}$= MJD 54265.1, \citealt[][]{CorbetKrimm2013:superorbital,Coley2015:eclipsing_binaries}).}
  \label{supero:fig:16493_phase_fits}
\end{figure}%%%%%%%%%%%%%%%%%%%%%%%%%%%%%%%%%%%%%%%%%%%%%%%%%%%%%%%%%

    %%%%%%%%%%%%%%%%%%%%%%%%%%%%%%%%%%%%%%%%%%%%%%%%%%%%%%%%% TABLE 3
\vspace{-0.3cm}
\begin{table}
 \tabcolsep 3pt
 \begin{center}
\caption{Results of the superorbital phase-resolved spectral analysis for the three sources \srca, \srcb, and \srcc.  
 \label{supero:tab:s_o_phase_specfits} }
 \begin{tabular}{ ccccr }
 %   \noalign{\smallskip}
 \hline
 %\noalign{\smallskip}
  Superorbital             & $N_{\rm H}$   & $\Gamma$ & F$_{\rm 0.3-10 keV}$                            &  Cstat                         \cr
   phase                     & $10^{22}$   &                     & $10^{-11}$                                         &      /d.o.f.    \cr 
                                 &  (cm$^{-2}$)  &                     & (erg\,cm$^{-2}$\,s$^{-1}$) &          \cr 
 \noalign{\smallskip} 
\hline   
\noalign{\smallskip}
J16418 & & & & \\
 \hline
0.000--0.185 & $7.2^{+1.1}_{-1.0}$ & $0.9^{+0.2}_{-0.2}$  & $6.1^{+0.4}_{-0.4}$   &     556.0/608\\
0.185--0.310 & $4.5^{+1.4}_{-1.2}$ & $0.7^{+0.4}_{-0.3}$  & $29.8^{+3.9}_{-3.4}$   &     269.3/327\\
0.310--0.450 & $6.4^{+2.0}_{-1.7}$ & $1.1^{+0.4}_{-0.4}$  & $4.0^{+0.6}_{-0.5}$   &     263.3/294\\
0.450--0.500 & $8.1^{+1.7}_{-1.5}$ & $0.6^{+0.3}_{-0.3}$  & $16.6^{+1.4}_{-1.3}$   &     435.5/514\\
0.500--0.660 & -- & -- & -- & -- \\ 
0.660--0.770 & $6.2^{+3.0}_{-2.4}$ & $0.5^{+0.6}_{-0.5}$  & $38.5^{+7.5}_{-6.4}$   &     163.5/181\\
0.770--0.910 & $7.5^{+2.3}_{-1.9}$ & $0.9^{+0.4}_{-0.4}$  & $1.8^{+0.2}_{-0.2}$   &     304.5/374\\
0.910--1.000 & $14.9^{+3.8}_{-3.2}$ & $0.5^{+0.3}_{-0.3}$  & $4.5^{+0.3}_{-0.3}$   &     551.0/596\\
 \noalign{\smallskip} 
\hline   
\noalign{\smallskip}
 J16479 & & & & \\ 
\hline
 0.000--0.165 & $11.8^{+2.1}_{-1.9}$ & $0.9^{+0.3}_{-0.3}$  & $4.4^{+0.3}_{-0.3}$   &     542.5/582\\
0.165--0.273 & $11.6^{+1.7}_{-1.5}$ & $1.1^{+0.2}_{-0.2}$  & $2.0^{+0.1}_{-0.1}$   &     576.6/624\\
0.273--0.340 & $10.6^{+1.6}_{-1.5}$ & $1.1^{+0.2}_{-0.2}$  & $3.6^{+0.2}_{-0.2}$   &     500.6/591\\
0.340--0.450 & $7.5^{+1.2}_{-1.1}$ & $1.0^{+0.2}_{-0.2}$  & $3.7^{+0.3}_{-0.2}$   &     596.2/593\\
0.450--0.553 & $11.3^{+1.8}_{-1.6}$ & $1.2^{+0.3}_{-0.3}$  & $2.2^{+0.2}_{-0.1}$   &     495.2/580\\
0.553--0.575 & $8.3^{+1.4}_{-1.3}$ & $0.7^{+0.2}_{-0.2}$  & $7.4^{+0.5}_{-0.5}$   &     533.4/585\\
0.575--0.740 & $8.4^{+1.5}_{-1.4}$ & $0.8^{+0.3}_{-0.2}$  & $1.1^{+0.1}_{-0.1}$   &     498.0/574 \\
0.740--1.000 &   $9.3^{+1.7}_{-1.6}$ & $1.0^{+0.3}_{-0.3}$  & $1.0^{+0.1}_{-0.1}$   &     494.0/572\\ 
\noalign{\smallskip}
 \hline
 \noalign{\smallskip}
J16493 & & & & \\ 
 \hline
0.000--0.125 & $13.0^{+5.1}_{-4.2}$ & $0.3^{+0.5}_{-0.5}$  & $2.9^{+0.4}_{-0.3}$   &     275.8/332\\
0.125--0.185 & $8.0^{+2.1}_{-1.8}$ & $0.5^{+0.3}_{-0.3}$  & $4.3^{+0.4}_{-0.4}$   &     338.7/416\\
0.185--0.375 & $11.5^{+2.9}_{-2.5}$ & $0.7^{+0.4}_{-0.4}$  & $2.2^{+0.2}_{-0.2}$   &     377.8/443\\
0.375--0.493 & $8.8^{+2.3}_{-2.0}$ & $0.7^{+0.4}_{-0.3}$  & $2.6^{+0.3}_{-0.2}$   &     357.0/437\\
0.493--0.685 & $11.6^{+3.1}_{-2.6}$ & $0.7^{+0.4}_{-0.4}$  & $1.7^{+0.2}_{-0.2}$   &     324.3/435\\
0.685--0.810 & $10.8^{+2.8}_{-2.4}$ & $0.6^{+0.4}_{-0.3}$  & $3.0^{+0.3}_{-0.3}$   &     378.7/420\\
0.810--0.877 & $7.2^{+2.0}_{-1.7}$ & $0.6^{+0.3}_{-0.3}$  & $3.6^{+0.4}_{-0.3}$   &     344.3/425\\
0.877--1.000 & $15.3^{+4.3}_{-3.7}$ & $0.8^{+0.5}_{-0.5}$  & $3.0^{+0.4}_{-0.3}$   &     277.5/361\\
\noalign{\smallskip}
 \hline
 \noalign{\smallskip}
\end{tabular}
 \end{center}
\end{table}

   %%%%%%%%%%%%%%%%%%%%%%%%%%%%%%%%%%%%%%%%%%%%%%%%%%%%%%%%%
   \section{Results and discussion} \label{supero:discussion}  
   %%%%%%%%%%%%%%%%%%%%%%%%%%%%%%%%%%%%%%%%%%%%%%%%%%%%%%%%%

\subsection{Orbital variability}
\label{sec:orbital}

The suitability of XRT monitoring campaigns to perform orbital variability studies was already widely 
demonstrated by our group in a recent paper for a number of wind-fed binaries 
\citep{Ferrigno2022:SGXRB}. Here we discuss the orbital variability only for the source 2S~0114+650, as 
for the remaining sources considered in the present paper the corresponding results were already 
reported previously by our group and also extensively analyzed in the literature \citep[see][and 
references therein]{Romano2009:sfxts_paperV,Romano2011:sfxts_paperVI,Romano2012:sfxts_16418}.

The left side of Fig.~\ref{supero:fig:2S0114_orbital_phase_fits_campaign} shows the variability of the 
spectral parameters measured in the case of 2S~0114+650 as a function of the system orbital phase. As in 
the case of previous objects, XRT was able to unveil also for this source an interesting variability 
pattern for the flux, the absorption column density, and the power-law photon index. The phase 0 is set 
at the periastron passage in such a way that our figure and the equivalent one realized with \swift/BAT 
data (15--50\,keV) in \citet[][see their Fig.~2]{CorbetKrimm2013:superorbital} are directly comparable. 
Both lightcurves display a minimum in flux at phase 0.7 that was already visible in the folded \rxte/PCA 
data\footnote{Note that the source light curve folded on the system orbital period reported in 
\citet{Farrell2008} uses the time of the flux minimum to define phase 0 and thus there is a shift of 
$\sim$0.3 in phase compared to our and \citet{CorbetKrimm2013:superorbital} light curves.} (2--12~keV 
range) reported by \citet[see their Fig.~13;][]{Farrell2008} and extensively investigated by 
\citet{Pradhan2015} to verify the hypothesis of an X-ray eclipse. The latter authors concluded that the 
properties of the flux minimum are not compatible with being a total obscuration of the NS by its 
companion. This conclusion was mainly driven by the fact that \suzaku\ data collected during the flux 
minimum could only reveal a modest increase of the local column density (a factor of $\sim3$) and 
equivalent width of the fluorescence iron line (if any at all), as well as a complete lack of evidence 
for a significant flattening of the spectral slope. This is at odds with what is commonly observed in 
eclipsing wind-fed X-ray binaries where both the absorption column density and equivalent width of the 
fluorescence iron emission line can increase by a factor of $\sim$10--100 (compared to the 
out-of-eclipse emission), and the power-law slope gets dramatically softer due to the fact that only the 
scattered X-ray emission is observed during the eclipse \citep[rather than the direct X-ray emission 
from the accreting source; see, e.g.][and references therein]{Falanga2015}. These authors thus suggested 
that the flux minimum is most likely due to the modulation of the mass accretion rate along the 
relatively highly eccentric orbit (see Sect.~\ref{supero:2s}). The sensitivity achieved thanks to our 
XRT monitoring campaign allows us to confirm this suggestion, since during the flux minimum we recorded 
a noticeable but still modest increase of $N_{\rm H}$ compared to the remaining orbital phases \citep[a 
factor of $\sim$3 as reported also by][]{Pradhan2015}. We can consider that the XRT measurement is truly 
representative of the characteristic absorption column density in the X-ray dip because this value is 
obtained as an average of many different orbital cycles.

The folded XRT lightcurve also presents two peaks in flux preceding and following the periastron 
passage. Although this has not been widely discussed in the literature, it is worth remarking that the 
peak in flux following the periastron is expected in case of eccentric or short-orbital period wind-fed 
SgXB due to the effect of both the stellar wind photoionization by the high-energy emission of the 
compact object and the relative velocity between the orbiting NS and the stellar wind. As a consequence 
of these effects, a second peak in flux is also expected slightly before the NS approaches the 
periastron again \citep[with the exact position in phase depending also on the poorly known physical 
conditions of the stellar wind, as the density, velocity, and ionization state; see][and references 
therein]{Bozzo2021}. A similar double peaked lightcurve is observed in the case of the SFXT endowed with 
the shortest orbital period of this class of objects, IGR\,J16479-4514 
\citep[][Fig.~1]{Sidoli2013:16479}.

Note that the double peaked structure of the folded XRT lightcurve is not immediately evident from the 
equivalent folded lightcurves in the higher energy domains, e.g. those obtained from \swift/BAT. This is 
not unexpected as the X-ray emission of wind-fed SgXBs above $\gtrsim$10~keV is dominated by the cut-off 
power-law spectral component. This component usually shows a much less pronounced variability along the 
orbital phase \citep[see also the discussion in][]{Farrell2008} and it is only marginally affected by 
changes in the absorption column density local to the source. The folded BAT lightcurves along the 
orbital phases of SgXBs thus display in most of the cases a modest (if any at all) variability, apart 
from the evident cases of X-ray eclipses \citep[see][for recent 
reviews]{Falanga2015,Coley2015:eclipsing_binaries}. Before the provision of our XRT results here, a very 
different profile of the X-ray emission from 2S~0114+650 in the soft versus hard X-ray domain could be 
well appreciated by looking at Fig.~2 of \citet{Pradhan2015}. Although the \suzaku/XIS data covering the 
soft energy domain ($\lesssim$10~keV) only spanned a limited portion of the orbital phase around the 
X-ray minimum, the much more prominent variability of these data compared to the super-imposed ones from 
\swift/BAT is particularly striking.

\subsection{Superorbital variability}
\label{sec:superorbital} 

Of more interest for the focus of this paper are the variations of the source spectral parameters as a 
function of the superorbital phase, as shown in the right side of 
Fig.~\ref{supero:fig:2S0114_orbital_phase_fits_campaign}. In the past, only \citet{Farrell2008} 
attempted a superorbital-phase-resolved spectral analysis of the X-ray emission from 2S~0114+650, but 
the outcomes of their analysis were partly hampered by the limited coverage of the \rxte/PCA data used, 
which sampled only two superorbital cycles. The PCA folded light curve showed a peak of the emission at 
phase 0.5, and only a marginally significant increase in the photon index was reported around phase 1.0. 
The data also suggested a possible increase of the absorption column density at the same phase, but the 
uncertainty associated with the measurement was far too large to draw a firm conclusion. The higher 
sensitivity and longer coverage of our XRT monitoring campaign (spanning six superorbital cycles) allows 
us to improve the measurements and investigate in more detail the properties of the source X-ray 
emission at different superorbital phases. The profile of the XRT folded light curve shows a sharp peak 
in the source flux at phase 0.5 and an intriguing pattern of variability of the $N_{\rm H}$, confirming 
a remarkable increase toward phase 0.8--1.0 (note that our figure and Fig.~16 of \citet{Farrell2008} are 
directly compatible as the reference time for phase 0 is the same in both cases). We could not detect 
any changes in the power-law photon index, which remains virtually constant at a value of 0.8 at most 
superorbital phases and show a modest decrease down to $\sim$0.6 around the superorbital phase 0.5. We 
tested that a fit to the eight superorbital phase bin spectra with a common photon index (introducing 
inter-calibration constants and with all other parameters free) yields a photon index of 
$0.795_{-0.055}^{+0.056}$, which is consistent with all individual fits apart from the two spectra 
extracted around phase 0.5. However, in these two cases, the deviation is only marginally 
($\lesssim$3~$\sigma$~c.l.) significant.

\citet{Farrell2008} suggested that the superorbital periodicity in 2S~0114+650 could be due to a 
periodic variation of the mass accretion rate, although these authors could not identify any specific 
mechanism(s) driving the variation. As discussed in Sect.~\ref{supero:intro}, there is a general 
convergence in believing that the CIR model provides the most solid ground to explain the superorbital 
variability of wind-fed X-ray binaries as 2S~0114+650. In this model, increases in the flux are 
associated with the accretion onto the NS of the denser \citep[by a factor of 2--3; see][and reference 
stherein]{Lobel2008} and (possibly) slower material of the CIR compared to the surrounding stellar wind. 
This interaction should also lead to increases in the local absorption column density, because the CIR 
is expected to intercept at some point the line of sight between the compact object and the observer. 
Our XRT measurements unveiled increases in the flux and absorption column density that are compatible 
with the factor 2--3 expected in the CIR model, although some geometrical effect shall be assumed in 
order to interpret the difference in phase between the rise in the flux and that in the $N_{\rm H}$. 
Compatibly with the CIR model, no dramatic variation is observed in the power-law photon index, as 
accretion is driven at any phase by the stellar wind and no spectral state changes are expected as those 
observed in disk-accreting systems \citep{Bozzo2016}. In view of the pioneering measurements we 
presented with XRT, it is difficult to obtain insights into the possibility suggested by \citet{Hu2017} 
that the source is not a pure wind-fed system but a temporary accretion disk is forming around the NS 
when the super-orbital variability is more pronounced. The physics of temporary accretion disk formation 
around NS in wind-fed binaries is not yet known, and neither are the expected spectral energy variations 
as a function of the accretion disk properties \citep[see, e.g., the discussion 
in][]{Romano2015:17544sb}. However, should a temporary accretion disk be driving the super-orbital 
modulation in 2S~0114+650, one would likely expect to see spectral state changes (i.e., changes in the 
power-law photon index) at different superorbital phases as those recorded from, e.g. Her\,X-1 and 
SMC\,X-1 \citep[see, e.g.,][and references therein]{Brumback2021:herx1last,Brumback2023}. At present, 
our XRT data do not show clear evidence to support this but deeper and more extended observations of the 
source are clearly needed to consolidate the reported findings \citep[to be carried out by exploiting, 
e.g., the combination of flexibility, soft X-ray coverage, and large effective area of NICER,][]{nicer}.

Concerning the remaining three sources considered in this paper, we note that in the case of \srca{}, 
XRT revealed some intriguing variability pattern for the source flux as a function of the superorbital 
phase (see Fig.~\ref{supero:fig:16418_phase_fits}). However, the coverage in phase is not complete and 
the S/N in the different available bins is far too low to claim the detection of any meaningful 
variation in either the power-law photon index or the absorption column density. We tested that merging 
more phase bins together to increase the S/N would not allow us to have a reasonable number of points to 
search for spectral variations at different superorbital phases. In the case of \srcb{}, the intriguing 
flux pattern is also accompanied by a similar HR variability trend. There is some indication in the data 
for possible variations especially of the absorption column density, but more data are needed in the 
different phase bins in order to decrease the uncertainties on the spectral parameters and draw a more 
firm conclusion. Among all analyzed sources, \srcc{} is the least interesting from a superorbital 
analysis point of view. XRT could only reveal modest changes in the flux and did not provide indications 
for possible variations in either the absorption column density or the power-law photon index. It should 
be noted, however, that the source is relatively faint for XRT and the error bars associated with both 
$N_{\rm H}$ and $\Gamma$ are definitively larger than in other cases (thus hampering any attempt to 
unveil only modest variations of these parameters). A substantially larger number of observations would 
be needed in this case to lower the error bars and dig into the presence of possibly modest (but 
recurrent) variations of the source spectral parameters as a function of the superorbital phase.

   %%%%%%%%%%%%%%%%%%%%%%%%%%%%%%%%%%%%%%%%%%%%%%%%%%%%%%%%%
   \section*{Data availability} \label{supero:data_archives}
   %%%%%%%%%%%%%%%%%%%%%%%%%%%%%%%%%%%%%%%%%%%%%%%%%%%%%%%%%
The data underlying this article are publicly available from the 
\swift\ archive and processed with publicly available software.

   %%%%%%%%%%%%%%%%%%%%%%%%%%%%%%%%%%%%%%%%%%%%%%%%%%%%%%%%%
    \section*{Acknowledgements} \label{supero:thanks}
   %%%%%%%%%%%%%%%%%%%%%%%%%%%%%%%%%%%%%%%%%%%%%%%%%%%%%%%%%
We acknowledge unwavering support from Amos. 
We also thank the anonymous referee for comments that helped   improve the paper. 
We acknowledge financial contribution from contracts ASI-INAF I/037/12/0 and ASI-INAF n.\ 2017-14-H.0.  
NI acknowledges support from NASA grants 80NSSC21K0022 and 80NSSC21K1994. 
This work was supported in part by NASA under award number 80GSFC21M0002. 
The \swift\ data of our monitoring campaigns on 2S~0114+650,  
on IGR~J16493$-$4348 and 
(partially) IGR~J16418$-$4532 were obtained through contract 
ASI-INAF I/004/11/0 to ASI-INAF I/004/11/5. 
This work made use of data supplied by the UK Swift Science Data Centre at the
University of Leicester \citep[see][]{Evans2007:repository,Evans2009:xrtgrb}. 
Happy 18th, {\it Swift}.

\bibliographystyle{mnras}

%\clearpage

   %%%%%%%%%%%%%%%%%%%%%%%%%%%%%%%%%%%%%%%%%%%%%%%%%%%%%%%%% APPENDICES  
   \appendix
   %%%%%%%%%%%%%%%%%%%%%%%%%%%%%%%%%%%%%%%%%%%%%%%%%%%%%%%%%

   %%%%%%%%%%%%%%%%%%%%%%%%%%%%%%%%%%%%%%%%%%%%%%%%%%%%%%%%%
    \section{Logs of observations} \label{supero:appendix1} 
   %%%%%%%%%%%%%%%%%%%%%%%%%%%%%%%%%%%%%%%%%%%%%%%%%%%%%%%%%
In this Section we include several tables to complement the results described in the main paper.  
Table~\ref{supero:tab:swift_xrt_log_s0114} reports all details of the new observational campaign 
focused on \mysrc{} (and never reported elsewhere), 
while Tables~\ref{supero:tab:swift_xrt_log_16418}, \ref{supero:tab:swift_xrt_log_16479},  
 and \ref{supero:tab:swift_xrt_log_16493} those of the campaigns on
\srca, \srcb, and \srcc.

              %%%%%%%%%%%%%%%%%%%%%%%%%%%%%%%%%%%%%%%%%%%%%%%%%%%%%%%%%% TABLE A1
              %\input{table_a1_2S0114_xrt_log_2phase.tex}  % observing log + orbital and superorbital phase + fluxes for each obs
              %%%%%%%%%%%%%%%%%%%%%%%%%%%%%%%%%%%%%%%%%%%%%%%%%%%%%%%%%% 
 \setcounter{table}{0}		
 \begin{table*} 	
 \begin{center} 
  \caption{Observation log for 2S~0114$+$650: ObsID, date (MJD of the middle of the observation), orbital phase, superorbital phase, start and end times (UTC), XRT exposure time, 
and  0.3--10\,keV observed and unabsorbed flux. \label{supero:tab:swift_xrt_log_s0114} } 	
  %\scriptsize % < ------------------------------------------ !!! 
    \begin{tabular}{cc ll ll rcc}
 \hline
 \noalign{\smallskip}
      ObsID   & MJD           &  Orbital$^a$        &  Super-o.$^b$   &    Start time (UT)                 &  End time (UT)             &  Exposure & Flux$^c$ & Flux$^d$ \\
                       &                   & Phase          &    Phase           &   (yyyy-mm-dd hh:mm:ss) & (yyyy-mm-dd hh:mm:ss)      &  (s)      & (erg\,cm$^{-2}$\,s$^{-1}$) & (erg\,cm$^{-2}$\,s$^{-1}$) \\
  \noalign{\smallskip}
 \hline
 \noalign{\smallskip}
00015874001 & 59985.97953 & 0.65 & 0.25 & 2023-02-10 23:23:09 & 2023-02-10 23:37:52 & 883  &	$2.5^{+0.8}_{-0.6}$ & $13.7^{+11.5}_{-8.6}$ \\ 
00015874002 & 59987.63298 & 0.80 & 0.30 & 2023-02-12 15:05:05 & 2023-02-12 15:17:52 & 767  &	$21.3^{+3.3}_{-2.8}$ & $28.4^{+4.4}_{-3.1}$ \\ 
00015874003 & 59989.36260 & 0.94 & 0.36 & 2023-02-14 08:34:24 & 2023-02-14 08:49:52 & 928  &	$1.2^{+0.4}_{-0.3}$ & $1.7^{+0.8}_{-0.4}$ \\ 
00015874004 & 59992.53420 & 0.22 & 0.46 & 2023-02-17 12:40:37 & 2023-02-17 12:57:52 & 1036  &	$55.8^{+5.8}_{-5.2}$ & $63.2^{+5.1}_{-4.7}$ \\ 
00015874005 & 59994.18332 & 0.36 & 0.51 & 2023-02-19 04:17:05 & 2023-02-19 04:30:52 & 827  &	$36.1^{+4.4}_{-3.9}$ & $43.4^{+4.0}_{-3.7}$ \\ 
00015874006 & 59996.17138 & 0.53 & 0.58 & 2023-02-21 03:59:42 & 2023-02-21 04:13:52 & 850  &	$21.3^{+4.0}_{-3.4}$ & $52.8^{+19.8}_{-16.8}$ \\ 
00015874007 & 59999.09258 & 0.78 & 0.67 & 2023-02-24 02:05:43 & 2023-02-24 02:20:53 & 910  &	$16.3^{+3.2}_{-2.6}$ & $22.7^{+6.3}_{-3.3}$ \\ 
00015874008 & 60001.14424 & 0.96 & 0.74 & 2023-02-26 03:20:31 & 2023-02-26 03:34:53 & 863  &	$2.4^{+0.6}_{-0.5}$ & $3.7^{+2.4}_{-0.8}$ \\ 
00015874009 & 60003.59002 & 0.17 & 0.82 & 2023-02-28 14:05:20 & 2023-02-28 14:13:54 & 514  &	$7.1^{+1.4}_{-1.2}$ & $9.0^{+1.6}_{-1.2}$ \\ 
00015874010 & 60006.39117 & 0.41 & 0.91 & 2023-03-03 00:34:40 & 2023-03-03 18:11:53 & 1860  &	$8.7^{+1.1}_{-1.0}$ & $12.9^{+2.4}_{-1.4}$ \\ 
00015874011 & 60009.07858 & 0.64 & 1.00 & 2023-03-05 06:44:26 & 2023-03-06 21:01:53 & 3229  &	 --  &	 --  \\ 
00015874012 & 60010.53555 & 0.77 & 0.05 & 2023-03-07 01:42:29 & 2023-03-07 23:59:53 & 3543  &	$0.2^{+0.1}_{-0.1}$ & $0.4^{+0.7}_{-0.1}$ \\ 
00015874013 & 60012.52010 & 0.94 & 0.11 & 2023-03-09 12:22:00 & 2023-03-09 12:35:52 & 832  &	$2.7^{+0.7}_{-0.6}$ & $3.6^{+0.9}_{-0.6}$ \\ 
00015874014 & 60015.09010 & 0.16 & 0.19 & 2023-03-12 02:01:35 & 2023-03-12 02:17:53 & 978  &	$21.2^{+2.7}_{-2.4}$ & $26.8^{+2.9}_{-2.5}$ \\ 
00015874015 & 60017.87524 & 0.40 & 0.28 & 2023-03-14 20:51:47 & 2023-03-14 21:08:53 & 1025  &	$29.0^{+3.8}_{-3.3}$ & $54.6^{+21.8}_{-9.7}$ \\ 
00015874016 & 60019.49996 & 0.54 & 0.34 & 2023-03-16 03:12:00 & 2023-03-16 20:47:53 & 1246  &	$1.1^{+0.6}_{-0.4}$ & $1.1^{+0.6}_{-0.4}$ \\ 
00015874017 & 60022.43528 & 0.80 & 0.43 & 2023-03-19 00:45:42 & 2023-03-19 20:07:53 & 1377  &	$13.9^{+1.7}_{-1.5}$ & $19.9^{+3.2}_{-2.1}$ \\ 
00015874018 & 60024.41987 & 0.97 & 0.50 & 2023-03-21 02:03:20 & 2023-03-21 18:05:52 & 1311  &	$42.8^{+3.9}_{-3.5}$ & $53.8^{+4.1}_{-3.6}$ \\ 
00015874019 & 60026.00975 & 0.10 & 0.55 & 2023-03-23 00:08:12 & 2023-03-23 00:19:52 & 700  &	$20.2^{+3.8}_{-3.1}$ & $36.1^{+20.0}_{-7.7}$ \\ 
00015874020 & 60035.92511 & 0.96 & 0.87 & 2023-04-01 21:20:25 & 2023-04-01 23:03:53 & 700  &	$6.0^{+1.2}_{-1.0}$ & $10.0^{+5.3}_{-2.0}$ \\
00015874021 & 60038.80610 & 0.21 & 0.97 & 2023-04-04 19:13:40 & 2023-04-04 19:27:53 & 852  &	$3.6^{+0.9}_{-0.7}$ & $3.8^{+0.8}_{-0.7}$ \\ 
00015874022 & 60040.56001 & 0.36 & 0.02 & 2023-04-06 06:21:57 & 2023-04-06 20:30:51 & 945  &	$17.7^{+2.9}_{-2.4}$ & $28.6^{+10.4}_{-4.5}$ \\ 
00015874023 & 60042.10699 & 0.49 & 0.07 & 2023-04-08 02:26:13 & 2023-04-08 02:41:53 & 940  &	$2.4^{+0.8}_{-0.6}$ & $3.8^{+4.9}_{-1.0}$ \\ 
00015874024 & 60045.58485 & 0.79 & 0.19 & 2023-04-11 08:17:29 & 2023-04-11 19:46:52 & 1138  &	$1.0^{+0.3}_{-0.2}$ & $1.7^{+1.9}_{-0.4}$ \\ 
00015874025 & 60047.20792 & 0.93 & 0.24 & 2023-04-13 04:51:53 & 2023-04-13 05:06:53 & 900  &	$0.6^{+0.4}_{-0.2}$ & $0.8^{+3.1}_{-0.3}$ \\ 
00015874026 & 60049.24266 & 0.11 & 0.30 & 2023-04-15 05:41:59 & 2023-04-15 05:56:51 & 893  &	$6.8^{+1.7}_{-1.3}$ & $7.7^{+1.4}_{-1.1}$ \\ 
00015874027 & 60052.28427 & 0.37 & 0.40 & 2023-04-18 06:41:48 & 2023-04-18 06:56:53 & 905  &	 --  &	 --  \\ 
00015874028 & 60054.60630 & 0.57 & 0.48 & 2023-04-20 14:25:16 & 2023-04-20 14:40:52 & 935  &	 --  &	 --  \\ 
00015874029 & 60056.85598 & 0.76 & 0.55 & 2023-04-22 20:24:18 & 2023-04-22 20:40:53 & 995  &	$1.7^{+0.4}_{-0.3}$ & $2.5^{+1.3}_{-0.5}$ \\ 
00015874030 & 60058.03684 & 0.87 & 0.59 & 2023-04-24 00:46:13 & 2023-04-24 00:59:53 & 820  &	$50.8^{+6.2}_{-5.5}$ & $63.5^{+6.1}_{-5.5}$ \\ 
00015874032 & 60063.07469 & 0.30 & 0.75 & 2023-04-29 01:40:14 & 2023-04-29 01:54:52 & 878  &	$11.5^{+2.4}_{-2.0}$ & $15.8^{+4.3}_{-2.3}$ \\ 
00015874033 & 60065.28811 & 0.49 & 0.83 & 2023-04-30 22:06:51 & 2023-05-01 15:42:53 & 1800  &	$9.8^{+1.2}_{-1.1}$ & $12.7^{+1.4}_{-1.2}$ \\ 
00015874034 & 60068.02361 & 0.73 & 0.91 & 2023-05-04 00:28:08 & 2023-05-04 00:39:52 & 705  &	$80.2^{+9.6}_{-8.5}$ & $117.9^{+20.3}_{-12.6}$ \\ 
00015874035 & 60070.06917 & 0.90 & 0.98 & 2023-05-06 01:32:19 & 2023-05-06 01:46:52 & 873  &	$27.1^{+4.2}_{-3.6}$ & $41.2^{+12.3}_{-5.8}$ \\ 
00015874036 & 60072.12550 & 0.08 & 0.05 & 2023-05-08 02:51:33 & 2023-05-08 03:09:53 & 1101  &	$28.0^{+4.0}_{-3.5}$ & $32.6^{+3.5}_{-3.3}$ \\ 
00015874037 & 60075.62487 & 0.38 & 0.16 & 2023-05-11 14:53:43 & 2023-05-11 15:05:53 & 730  &	$1.6^{+0.5}_{-0.4}$ & $3.2^{+4.5}_{-0.9}$ \\ 
00015874038 & 60077.01463 & 0.50 & 0.21 & 2023-05-13 00:14:15 & 2023-05-13 00:27:51 & 767  &	$0.7^{+1.6}_{-0.5}$ & $0.8^{+1.5}_{-0.3}$ \\ 
00015874039 & 60088.94118 & 0.53 & 0.59 & 2023-05-24 22:31:42 & 2023-05-24 22:38:53 & 431  &	$1.1^{+1.2}_{-0.6}$ & $2.1^{+1.3}_{-0.9}$ \\ 
00015874040 & 60091.72302 & 0.77 & 0.69 & 2023-05-27 17:14:23 & 2023-05-27 17:27:53 & 810  &	$0.7^{+0.3}_{-0.2}$ & $2.1^{+1.6}_{-1.1}$ \\ 
00015874041 & 60093.62734 & 0.93 & 0.75 & 2023-05-29 14:55:49 & 2023-05-29 15:10:54 & 905  &	$7.8^{+1.7}_{-1.4}$ & $9.2^{+1.5}_{-1.3}$ \\ 
00015874042 & 60095.80837 & 0.12 & 0.82 & 2023-05-31 19:16:13 & 2023-05-31 19:31:53 & 940  &	$32.8^{+3.8}_{-3.4}$ & $35.3^{+3.4}_{-3.0}$ \\ 
00015874043 & 60097.33644 & 0.25 & 0.87 & 2023-06-02 03:13:04 & 2023-06-02 12:55:53 & 1344  &	$9.0^{+1.1}_{-1.0}$ & $12.6^{+1.8}_{-1.2}$ \\ 
00015874044 & 60100.11498 & 0.49 & 0.96 & 2023-06-05 01:01:14 & 2023-06-05 04:29:53 & 1710  &	$1.0^{+0.4}_{-0.3}$ & $1.2^{+0.8}_{-0.3}$ \\ 
00015874045 & 60102.18849 & 0.67 & 0.03 & 2023-06-07 03:40:56 & 2023-06-07 05:21:53 & 1387  &	 --  &	 --  \\ % 
00015874046 & 60104.14462 & 0.84 & 0.09 & 2023-06-09 01:43:38 & 2023-06-09 05:11:52 & 1519  &	$15.6^{+2.1}_{-1.9}$ & $18.3^{+1.8}_{-1.7}$ \\ 
00015874047 & 60107.02621 & 0.09 & 0.18 & 2023-06-11 22:12:35 & 2023-06-12 03:02:53 & 1066  &	$7.5^{+1.4}_{-1.2}$ & $10.2^{+2.4}_{-1.4}$ \\ 
00015874048 & 60109.09067 & 0.27 & 0.25 & 2023-06-14 02:03:14 & 2023-06-14 02:17:54 & 880  &	$18.2^{+2.6}_{-2.2}$ & $26.3^{+5.1}_{-3.1}$ \\ 
00015874049 & 60110.71524 & 0.41 & 0.30 & 2023-06-15 10:02:59 & 2023-06-16 00:16:53 & 875  &	$3.0^{+0.8}_{-0.6}$ & $5.5^{+6.1}_{-1.4}$ \\ 
00015874050 & 60114.32445 & 0.72 & 0.42 & 2023-06-19 07:39:32 & 2023-06-19 07:54:52 & 920  &	$3.0^{+0.7}_{-0.5}$ & $5.8^{+5.4}_{-1.6}$ \\ 
00015874051 & 60116.23916 & 0.88 & 0.48 & 2023-06-21 05:34:52 & 2023-06-21 05:53:53 & 1141  &	$22.2^{+3.7}_{-3.2}$ & $25.0^{+3.3}_{-2.9}$ \\ 
00015874052 & 60118.93825 & 0.12 & 0.57 & 2023-06-23 22:23:15 & 2023-06-23 22:38:53 & 938  &	$1.1^{+0.4}_{-0.3}$ & $2.4^{+5.5}_{-0.8}$ \\ 
00015874053 & 60121.26255 & 0.32 & 0.65 & 2023-06-26 06:10:15 & 2023-06-26 06:25:53 & 938  &	 --  &	 --  \\ 
00015874054 & 60123.87309 & 0.54 & 0.73 & 2023-06-28 20:07:38 & 2023-06-28 21:46:52 & 1118  &  --  &	 --  \\ 
00015874055 & 60125.15684 & 0.65 & 0.77 & 2023-06-30 03:38:49 & 2023-06-30 03:52:52 & 842  &	 --  &	 --  \\ 
 \noalign{\smallskip}
  \hline
  \end{tabular}
  \end{center}
  \end{table*} 

\setcounter{table}{0}		
 \begin{table*} 	
 \begin{center} 
  \caption{Observation log for 2S~0114$+$650, continued.}	
  \begin{tabular}{cc ll ll rcc}
 \hline
 \noalign{\smallskip}
      ObsID   & MJD           &  Orbital$^a$        &  Super-o.$^b$   &    Start time (UT)                 &  End time (UT)             &  Exposure & Flux$^c$ & Flux$^d$ \\
                       &                   & Phase          &    Phase           &   (yyyy-mm-dd hh:mm:ss) & (yyyy-mm-dd hh:mm:ss)      &  (s)      & (erg\,cm$^{-2}$\,s$^{-1}$) & (erg\,cm$^{-2}$\,s$^{-1}$) \\
  \noalign{\smallskip}
 \hline
 \noalign{\smallskip}
00015874056 & 60127.59959 & 0.86 & 0.85 & 2023-07-02 14:14:56 & 2023-07-02 14:31:52 & 1015  &	$2.1^{+0.6}_{-0.5}$ & $2.8^{+0.7}_{-0.5}$ \\ 
00015874057 & 60129.54923 & 0.03 & 0.92 & 2023-07-04 01:13:52 & 2023-07-05 01:07:54 & 1101  &	$1.3^{+0.4}_{-0.3}$ & $1.7^{+0.6}_{-0.3}$ \\ 
00015874058 & 60141.69904 & 0.08 & 0.31 & 2023-07-16 16:38:20 & 2023-07-16 16:54:53 & 993  &	 --  &	 --     \\ % 
00015874059 & 60144.53360 & 0.32 & 0.40 & 2023-07-19 12:40:54 & 2023-07-19 12:55:51 & 898  &	$29.4^{+4.1}_{-3.6}$ & $39.2^{+5.5}_{-4.1}$ \\ 
00015874060 & 60146.46352 & 0.49 & 0.47 & 2023-07-21 01:33:04 & 2023-07-21 20:41:52 & 1038  &	$6.5^{+1.7}_{-1.4}$ & $8.3^{+2.8}_{-1.5}$ \\ 
00015874061 & 60148.17633 & 0.64 & 0.52 & 2023-07-23 02:33:57 & 2023-07-23 05:53:52 & 1249  &	$4.4^{+0.9}_{-0.7}$ & $7.9^{+6.9}_{-1.8}$ \\ 
00015874062 & 60150.41982 & 0.83 & 0.59 & 2023-07-25 08:33:12 & 2023-07-25 11:35:52 & 1111  &	$20.3^{+2.4}_{-2.1}$ & $26.4^{+2.8}_{-2.3}$ \\ 
00015874063 & 60153.92619 & 0.13 & 0.71 & 2023-07-28 22:05:32 & 2023-07-28 22:21:52 & 980  &	$8.2^{+1.6}_{-1.3}$ & $13.9^{+7.9}_{-2.8}$ \\ 
00015874064 & 60155.90796 & 0.30 & 0.77 & 2023-07-30 21:40:02 & 2023-07-30 21:54:52 & 890  &	--  &	 --     \\ % 
00015874065 & 60157.35743 & 0.43 & 0.82 & 2023-08-01 05:28:30 & 2023-08-01 11:40:52 & 955  &	 --  &	 --     \\ % 
00015874066 & 60160.59398 & 0.71 & 0.92 & 2023-08-04 14:07:46 & 2023-08-04 14:22:54 & 908  &	$1.8^{+0.5}_{-0.4}$ & $3.5^{+6.9}_{-1.1}$ \\ 
00015874067 & 60162.60713 & 0.88 & 0.99 & 2023-08-06 13:48:38 & 2023-08-06 15:19:52 & 1178  &	$4.2^{+0.9}_{-0.7}$ & $5.9^{+1.6}_{-0.9}$ \\ 
00015874068 & 60164.22094 & 0.02 & 0.04 & 2023-08-08 05:10:41 & 2023-08-08 05:25:54 & 895  &	$3.9^{+0.7}_{-0.6}$ & $6.2^{+2.2}_{-1.1}$ \\ 
00015874069 & 60168.04810 & 0.35 & 0.17 & 2023-08-11 12:22:37 & 2023-08-12 13:55:53 & 1108  &	$2.9^{+0.6}_{-0.5}$ & $4.4^{+2.2}_{-0.8}$ \\ 
00015874070 & 60170.00772 & 0.52 & 0.23 & 2023-08-13 17:00:19 & 2023-08-14 07:21:53 & 1364  &	$0.5^{+0.3}_{-0.2}$ & $8.5^{+11.4}_{-7.6}$ \\ 
00015874071 & 60171.55790 & 0.65 & 0.28 & 2023-08-15 11:46:51 & 2023-08-15 14:59:54 & 1046  &	$13.9^{+2.3}_{-1.9}$ & $35.1^{+12.8}_{-10.6}$ \\ 
00015874072 & 60173.20319 & 0.80 & 0.33 & 2023-08-17 04:42:19 & 2023-08-17 05:02:52 & 1234  &	$17.0^{+2.7}_{-2.3}$ & $22.9^{+4.2}_{-2.7}$ \\ 
00015874073 & 60176.64572 & 0.09 & 0.45 & 2023-08-20 15:22:47 & 2023-08-20 15:36:52 & 845  &	$1.4^{+0.8}_{-0.5}$ & $1.5^{+0.6}_{-0.4}$ \\ 
00015874074 & 60178.62604 & 0.26 & 0.51 & 2023-08-22 14:54:07 & 2023-08-22 15:08:52 & 885  &	$2.6^{+0.8}_{-0.6}$ & $3.2^{+1.0}_{-0.6}$ \\ 
00015874075 & 60180.27359 & 0.40 & 0.56 & 2023-08-24 06:27:03 & 2023-08-24 06:40:53 & 830  &	$3.9^{+1.0}_{-0.8}$ & $4.9^{+1.1}_{-0.8}$ \\ 
 \noalign{\smallskip}
  \hline
  \end{tabular}
% do not remove empty line below

Notes: $^a$ Phase zero for the orbital period is the time of periastron passage at MJD 51824.8 \citep[][]{Grundstrom2007}.
$^b$ Phase zero for the superorbital period is the minimum of the folded light curve at MJD 53488  \citep[][]{Farrell2008}. 
$^c$ Observed flux in the 0.3--10\,keV energy band in units of $\times 10^{-11}$ erg\,cm$^{-2}$\,s$^{-1}$. 
When no value is provided, the observation yielded insufficient counts to perform spectral analysis. 
$^d$ Unabsorbed flux in the 0.3--10\,keV energy band in units of $\times 10^{-11}$ erg\,cm$^{-2}$\,s$^{-1}$. 
\end{center}
\end{table*}

              %%%%%%%%%%%%%%%%%%%%%%%%%%%%%%%%%%%%%%%%%%%%%%%%%%%%%%%%%% TABLE A2
  \setcounter{table}{1}		
 \begin{table*} 	
 \tabcolsep 4pt         
 \begin{center} 	
 \caption{Observation log for IGR~J16418$-$4532. 
Observing sequence, date (MJD of the middle of the observation), superorbital phase, start and end times (UTC), XRT exposure time, and  0.3--10\,keV observed and unabsorbed flux, and references. 
 \label{supero:tab:swift_xrt_log_16418}}	
 \begin{tabular}{ l l l ll r rr l } 
 \hline 
 \hline 
 \noalign{\smallskip} 

 ObsID           &  MJD   & Phase & Start time  (UT)                     & End time   (UT)                 & Exposure  & Flux$^a$ & Flux$^b$& Ref.   \\ 
                      &           &          &  (yyyy-mm-dd hh:mm:ss)  & (yyyy-mm-dd hh:mm:ss)  &(s)              & (erg\,cm$^{-2}$\,s$^{-1}$) & (erg\,cm$^{-2}$\,s$^{-1}$) &              \\
  \noalign{\smallskip} 
 \hline 
 \noalign{\smallskip} 
00031929001 & 55610.11915 & 0.90 & 2011-02-18 01:59:42 & 2011-02-18 03:43:25 & 1953 & $16.4^{+2.3}_{-2.0}$ & $30.0^{+14.4}_{-5.6}$ & 	\getrefnumber{tabRomano2012:sfxts_16418}	 \\ 
00031929002 & 55624.10599 & 0.85 & 2011-03-04 00:03:18 & 2011-03-04 05:01:56 & 2021 & $10.1^{+1.4}_{-1.2}$ & $13.5^{+1.8}_{-1.4}$ & 		 \\ 
00031929003 & 55755.51071 & 0.77 & 2011-07-13 09:46:54 & 2011-07-13 14:43:56 & 4756 & $6.9^{+0.6}_{-0.5}$ & $10.0^{+1.0}_{-0.8}$ & 		 \\ 
00031929004 & 55756.67068 & 0.85 & 2011-07-14 12:39:38 & 2011-07-14 19:31:55 & 4832 & $0.5^{+0.3}_{-0.2}$ & $4.4^{+31.9}_{-3.5}$ & 		 \\ 
00031929005 & 55757.61961 & 0.91 & 2011-07-15 06:42:49 & 2011-07-15 23:01:56 & 1081 & $1.4^{+0.8}_{-0.5}$ & $1.8^{+8.0}_{-0.3}$ & 		 \\ 
00031929006 & 55758.69515 & 0.98 & 2011-07-16 10:15:04 & 2011-07-16 23:06:57 & 396 & $7.1^{+3.7}_{-2.4}$ & $9.5^{+46.6}_{-2.5}$ & 		 \\ 
00031929007 & 55761.52284 & 0.18 & 2011-07-19 01:52:40 & 2011-07-19 23:19:56 & 2673 & $7.2^{+0.9}_{-0.8}$ & $9.8^{+1.3}_{-1.0}$ & 		 \\ 
00032037001 & 55760.96527 & 0.14 & 2011-07-18 21:28:27 & 2011-07-19 00:51:30 & 1941 &      --  &        --     	 & 		 \\ % 
00032037002 & 55762.43113 & 0.24 & 2011-07-20 05:18:41 & 2011-07-20 15:22:57 & 3503 & $5.6^{+0.6}_{-0.5}$ & $8.9^{+1.7}_{-1.0}$ & 		 \\ 
00032037003 & 55763.53270 & 0.31 & 2011-07-21 10:06:17 & 2011-07-21 15:27:53 & 3779 & $1.9^{+0.3}_{-0.3}$ & $2.7^{+0.7}_{-0.4}$ & 		 \\ 
00032037004 & 55764.70995 & 0.39 & 2011-07-22 15:19:42 & 2011-07-22 18:44:56 & 2818 &      --  &        --     	 & 		 \\ % 
00032037005 & 55765.56675 & 0.45 & 2011-07-23 11:48:18 & 2011-07-23 15:23:55 & 4145 & $3.7^{+0.4}_{-0.4}$ & $5.7^{+1.2}_{-0.7}$ & 		 \\ 
00032037006 & 55771.74599 & 0.87 & 2011-07-29 12:13:29 & 2011-07-29 23:34:56 & 4669 &      -- &       --     	 & 		 \\ 
00032037007 & 55772.24797 & 0.90 & 2011-07-30 01:01:29 & 2011-07-30 10:52:57 & 5082 & $0.5^{+0.1}_{-0.1}$ & $0.8^{+0.6}_{-0.2}$ & 		 \\ 
 \noalign{\smallskip} 
00523489000 & 56081.80123 & 0.92 & 2012-06-03 18:16:39 & 2012-06-03 20:10:52 & 2771 & $8.4^{+1.0}_{-0.9}$ & $66.2^{+192.8}_{-35.4}$ &  \getrefnumber{tabRomano2012:atel4148}, \getrefnumber{tabRomano2013:MI50x_sfxts}, \getrefnumber{tabRomano2023:sfxts_catII} 	 \\ 
00523489001 & 56082.65221 & 0.98 & 2012-06-04 13:06:06 & 2012-06-04 18:12:15 & 3280 & $4.2^{+0.6}_{-0.6}$ & $7.0^{+3.5}_{-1.2}$ & 		 \\ 
 \noalign{\smallskip} 
00552677000 & 56384.54853 & 0.47 & 2013-04-02 12:16:51 & 2013-04-02 14:02:54 & 2191 & $35.5^{+3.0}_{-2.7}$ & $55.5^{+7.5}_{-5.0}$ & 	\getrefnumber{tabRomano2023:sfxts_catII} 	 \\ 
\noalign{\smallskip} 
00571067000 & 56552.67883 & 0.89 & 2013-09-17 16:08:16 & 2013-09-17 16:26:44 & 1108 & $3.8^{+0.9}_{-0.7}$ & $7.5^{+20.5}_{-2.4}$ & \getrefnumber{tabRomano2023:sfxts_catII}		 \\ 
00032037008 & 56553.43540 & 0.94 & 2013-09-18 09:21:35 & 2013-09-18 11:32:22 & 975 &      --  &        --     	 & 		 \\ % 
00032037009 & 56554.24883 & 0.99 & 2013-09-19 00:18:42 & 2013-09-19 11:37:54 & 973 & $7.9^{+1.5}_{-1.3}$ & $11.7^{+3.9}_{-1.7}$ & 		 \\ 
00032037010 & 56555.07223 & 0.05 & 2013-09-20 01:36:05 & 2013-09-20 01:51:55 & 950 & $9.6^{+2.4}_{-1.9}$ & $17.1^{+20.0}_{-4.3}$ & 		 \\ 
00032037011 & 56556.26644 & 0.13 & 2013-09-21 06:15:24 & 2013-09-21 06:31:55 & 990 & $7.4^{+1.5}_{-1.2}$ & $16.6^{+18.5}_{-4.7}$ & 		 \\ 
00032037012 & 56557.61314 & 0.22 & 2013-09-22 14:34:56 & 2013-09-22 14:50:54 & 958 & $0.7^{+0.4}_{-0.3}$ & $1.4^{+15.8}_{-0.6}$ & 		 \\ 
00032037013 & 56558.40401 & 0.28 & 2013-09-23 09:33:38 & 2013-09-23 09:49:53 & 975 & $0.7^{+0.9}_{-0.4}$ & $0.8^{+1.0}_{-0.3}$ & 		 \\ 
00032037014 & 56559.68368 & 0.36 & 2013-09-24 16:16:06 & 2013-09-24 16:32:53 & 1008 & $0.6^{+0.3}_{-0.2}$ & $1.2^{+7.2}_{-0.5}$ & 		 \\ 
 \noalign{\smallskip} 
00032037015 & 56907.15963 & 0.95 & 2014-09-07 03:41:49 & 2014-09-07 03:57:54 & 965 & $4.7^{+1.2}_{-1.0}$ & $6.0^{+2.7}_{-1.1}$  	 & \getrefnumber{tabThisWork1}	 \\ 
00032037016 & 56910.69873 & 0.19 & 2014-09-10 13:16:25 & 2014-09-10 20:15:54 & 1088 & $6.8^{+1.7}_{-1.4}$ & $10.9^{+9.0}_{-2.2}$  	 & 	 \\ 
00032037017 & 56914.95235 & 0.48 & 2014-09-14 22:43:50 & 2014-09-14 22:58:55 & 905 & $1.6^{+0.7}_{-0.5}$ & $2.0^{+1.3}_{-0.5}$  	 & 	 \\ 
00032037018 & 56917.09152 & 0.63 & 2014-09-17 02:03:38 & 2014-09-17 02:19:55 & 978  &      --  &        --     	 & 	 \\ % 
00032037019 & 56921.94654 & 0.96 & 2014-09-21 22:35:06 & 2014-09-21 22:50:54 & 948 & $2.3^{+0.9}_{-0.6}$ & $3.3^{+3.3}_{-0.8}$  	 & 	 \\ 
00032037020 & 56924.61884 & 0.14 & 2014-09-24 14:48:50 & 2014-09-24 14:53:25 & 276 & $9.7^{+3.6}_{-2.5}$ & $19.1^{+66.2}_{-6.7}$  	 & 	 \\ 
00032037021 & 56928.13989 & 0.38 & 2014-09-28 03:20:40 & 2014-09-28 03:22:11 & 90  &      --  &        --     	 & 	 \\ % 
00032037022 & 56931.14320 & 0.58 & 2014-10-01 03:18:28 & 2014-10-01 03:33:55 & 928  &      --  &        --     	 & 	 \\ % 
00032037023 & 56935.01289 & 0.84 & 2014-10-05 00:10:10 & 2014-10-05 00:26:56 & 1005  &      --  &        --     	 & 	 \\ % 
00032037024 & 56938.21281 & 0.06 & 2014-10-08 04:58:59 & 2014-10-08 05:13:54 & 895  &      --  &        --     	 & 	 \\ % 
00032037025 & 56942.94418 & 0.38 & 2014-10-12 22:29:20 & 2014-10-12 22:49:54 & 1234 & $9.7^{+1.8}_{-1.5}$ & $15.0^{+6.6}_{-2.3}$  	 & 	 \\ 
00032037026 & 56945.71201 & 0.57 & 2014-10-15 14:43:38 & 2014-10-15 19:26:56 & 1031  &      --  &        --     	 & 	 \\ % 
00032037027 & 56949.53568 & 0.83 & 2014-10-19 09:40:03 & 2014-10-19 16:02:54 & 1221  &      --  &        --     	 & 	 \\ % 
00032037029 & 56955.07324 & 0.21 & 2014-10-25 01:34:01 & 2014-10-25 01:56:55 & 1374 & $1.6^{+0.7}_{-0.5}$ & $4.1^{+29.9}_{-1.7}$  	 & 	 \\ 
00032037030 & 57040.49987 & 0.00 & 2015-01-18 11:51:43 & 2015-01-18 12:07:54 & 970 & $9.3^{+1.7}_{-1.4}$ & $16.3^{+8.3}_{-3.1}$  	 & 	 \\ 
00032037031 & 57043.16719 & 0.19 & 2015-01-21 03:52:36 & 2015-01-21 04:08:54 & 978  &      --  &        --     	 & 	 \\ % 
00032037032 & 57047.68975 & 0.49 & 2015-01-25 16:25:33 & 2015-01-25 16:40:55 & 923 & $8.0^{+1.5}_{-1.3}$ & $11.1^{+2.4}_{-1.4}$  	 & 	 \\ 
00032037033 & 57050.95139 & 0.71 & 2015-01-28 22:42:06 & 2015-01-28 22:57:54 & 948   &      --  &        --     	 &  	 \\ 
00032037034 & 57054.02305 & 0.92 & 2015-02-01 00:24:29 & 2015-02-01 00:41:54 & 1046  &      --  &        --     	 & 	 \\ 
  \noalign{\smallskip} 
00639199000 & 57139.60458 & 0.73 & 2015-04-27 14:25:53 & 2015-04-27 14:35:17 & 564 & $38.5^{+7.5}_{-6.4}$ & $50.0^{+9.4}_{-6.7}$  	 & 	\getrefnumber{tabRomano2023:sfxts_catII}   \\ 
00032037036 & 57141.19595 & 0.84 & 2015-04-29 04:34:23 & 2015-04-29 04:49:56 & 933 & $3.4^{+0.7}_{-0.6}$ & $7.0^{+6.7}_{-1.8}$  	 & 	 \\ 
00032037037 & 57142.98612 & 0.96 & 2015-04-30 23:33:21 & 2015-04-30 23:46:39 & 797 & $1.1^{+0.5}_{-0.3}$ & $16.9^{+179.5}_{-14.9}$   & 	 \\ 
00032037038 & 57143.24946 & 0.98 & 2015-05-01 01:10:31 & 2015-05-01 10:47:55 & 707 & $3.1^{+1.2}_{-0.8}$ & $7.0^{+81.0}_{-2.8}$  	 & \getrefnumber{tabRomano2023:sfxts_catII}	 \\ 
00032037039 & 57144.85102 & 0.09 & 2015-05-02 20:18:02 & 2015-05-02 20:32:54 & 893 & $4.3^{+1.3}_{-1.0}$ & $6.1^{+3.2}_{-1.1}$  	 & 	 \\ 
\noalign{\smallskip} 
01073821000 & 59475.38643 & 0.31 & 2021-09-18 09:08:42 & 2021-09-18 09:24:13 & 930 & $30.5^{+3.9}_{-3.5}$ & $39.3^{+4.4}_{-3.7}$  	 & \getrefnumber{tabSbarufatti2021:atel14924}, \getrefnumber{tabRomano2023:sfxts_catII} 	 \\ 
01073821001 & 59475.48760 & 0.31 & 2021-09-18 10:43:44 & 2021-09-18 12:40:33 & 2951 & $4.0^{+0.6}_{-0.5}$ & $6.3^{+1.6}_{-0.9}$  		 \\ 
\noalign{\smallskip} 
 \hline
  \end{tabular}
  \end{center}
  \end{table*}

\setcounter{table}{1}		
 \begin{table*} 	
\tabcolsep 4pt         
   \begin{center}
  \caption{Observation log for IGR~J16418$-$4532. Continued.} 
 \begin{tabular}{ l l l ll r rr l } 
 \hline 
 \hline 
 \noalign{\smallskip} 

 ObsID           &  MJD   & Phase & Start time  (UT)                     & End time   (UT)                 & Exposure  & Flux$^a$ & Flux$^b$& Ref.   \\ 
                      &           &          &  (yyyy-mm-dd hh:mm:ss)  & (yyyy-mm-dd hh:mm:ss)  &(s)              & (erg\,cm$^{-2}$\,s$^{-1}$) & (erg\,cm$^{-2}$\,s$^{-1}$) &              \\
  \noalign{\smallskip} 
 \hline 
 \noalign{\smallskip} 
00043105001	&	55739.22181	&	0.66	&	2011-06-27 05:14:53	&	2011-06-27 05:23:55	&	542  & $3.5^{+1.8}_{-1.2}$ & $4.6^{+4.1}_{-1.2}$ &   \getrefnumber{tabThisWork2} \\ 
00081650001	&	57291.94305	&	0.07	&	2015-09-26 21:43:06	&	2015-09-26 23:32:53	&	1891  & $6.2^{+1.0}_{-0.9}$ & $7.8^{+1.1}_{-0.9}$   & \\ 
\noalign{\smallskip}
00088748001	&	58393.93698	&	0.89	&	2018-10-02 21:32:47	&	2018-10-02 23:25:52	&	1800  & $2.0^{+0.4}_{-0.4}$ & $2.6^{+0.8}_{-0.4}$   &  \getrefnumber{tabIslam2023_superorbital} \\ 
00088748002	&	58401.90480	&	0.43	&	2018-10-10 20:45:56	&	2018-10-10 22:39:52	&	1790  & $4.4^{+0.7}_{-0.6}$ & $6.0^{+1.3}_{-0.8}$   &    \\ 
  \noalign{\smallskip} 
  \hline
  \end{tabular}

Notes: $^a$ Observed flux in the 0.3--10\,keV energy band in units of $\times 10^{-11}$ erg\,cm$^{-2}$\,s$^{-1}$.
When no value is provided, the observation yielded insufficient counts to perform spectral analysis.
$^b$ Unabsorbed flux in the 0.3--10\,keV energy band in units of $\times 10^{-11}$ erg\,cm$^{-2}$\,s$^{-1}$.
  \end{center}
\newcounter{ctr_so16418_tabrefs} 
\setrefcountdefault{-99} 
\refstepcounter{ctr_so16418_tabrefs}\label{tabRomano2012:sfxts_16418}(\getrefnumber{tabRomano2012:sfxts_16418}) Monitoring campaign in \citet{Romano2012:sfxts_16418}; 
\refstepcounter{ctr_so16418_tabrefs}\label{tabRomano2012:atel4148}(\getrefnumber{tabRomano2012:atel4148}) \citet{Romano2012:atel4148}; 
\refstepcounter{ctr_so16418_tabrefs}\label{tabRomano2013:MI50x_sfxts}(\getrefnumber{tabRomano2013:MI50x_sfxts}) \citet{Romano2013:MI50x_sfxts}; 
\refstepcounter{ctr_so16418_tabrefs}\label{tabRomano2023:sfxts_catII}(\getrefnumber{tabRomano2023:sfxts_catII}) Outbursts and followups in \citet{Romano2023:sfxts_catII}; 
\refstepcounter{ctr_so16418_tabrefs}\label{tabThisWork1}(\getrefnumber{tabThisWork1}) Monitoring campaign, this work.
\refstepcounter{ctr_so16418_tabrefs}\label{tabSbarufatti2021:atel14924}(\getrefnumber{tabSbarufatti2021:atel14924}) \citet{Sbarufatti2021:atel14924}; 
\refstepcounter{ctr_so16418_tabrefs}\label{tabThisWork2}(\getrefnumber{tabThisWork2}) Serendipitous observations, this work;
\refstepcounter{ctr_so16418_tabrefs}\label{tabIslam2023_superorbital}(\getrefnumber{tabIslam2023_superorbital}) \citet{Islam2023_superorbital}. 
 \end{table*}

              %%%%%%%%%%%%%%%%%%%%%%%%%%%%%%%%%%%%%%%%%%%%%%%%%%%%%%%%%% TABLE A3
 \setcounter{table}{2}
 \begin{table*}
\tabcolsep 4pt
   \begin{center}
  \caption{Observation log  for IGR~J16479$-$4514. 
Observing sequence, date (MJD of the middle of the observation), superorbital phase, start and end times (UTC), XRT exposure time, and  0.3--10\,keV observed and unabsorbed flux. 
 \label{supero:tab:swift_xrt_log_16479}}
  \begin{tabular}{ l l l ll r rr  }
%\hline
 \hline
% \noalign{\smallskip}
 ObsID           &  MJD   & Phase & Start time  (UT)                     & End time   (UT)                 & Exposure  & Flux & Flux      \\
                      &           &          &  (yyyy-mm-dd hh:mm:ss)  & (yyyy-mm-dd hh:mm:ss)  &(s)              & (erg\,cm$^{-2}$\,s$^{-1}$) & (erg\,cm$^{-2}$\,s$^{-1}$)                \\
 \noalign{\smallskip}
 \hline
\noalign{\smallskip}
00030296006 & 54484.96264 & 0.81 & 2008-01-19 22:58:28 & 2008-01-19 23:13:58 & 928 & $12.5^{+3.3}_{-2.6}$ & $15.9^{+3.6}_{-2.6}$  	 \\
00030296007 & 54487.84512 & 0.05 & 2008-01-22 20:07:57 & 2008-01-22 20:25:58 & 1081 & $2.5^{+0.6}_{-0.5}$ & $4.6^{+5.7}_{-1.2}$  	 \\ 
00030296008 & 54491.79550 & 0.38 & 2008-01-26 18:56:04 & 2008-01-26 19:14:57 & 1133 & $1.4^{+0.6}_{-0.4}$ & $2.2^{+3.2}_{-0.5}$  	 \\ 
00030296009 & 54494.34173 & 0.60 & 2008-01-29 08:03:13 & 2008-01-29 08:20:56 & 1063 & $0.4^{+0.4}_{-0.2}$ & $0.5^{+0.6}_{-0.2}$  	 \\ 
00030296010 & 54501.37081 & 0.19 & 2008-02-05 00:48:59 & 2008-02-05 16:58:56 & 3056 & $1.8^{+0.3}_{-0.2}$ & $8.3^{+21.8}_{-4.1}$  	 \\ 
00030296011 & 54505.89050 & 0.57 & 2008-02-09 18:53:39 & 2008-02-09 23:50:58 & 2751 & $4.2^{+0.6}_{-0.5}$ & $5.2^{+0.5}_{-0.5}$  	 \\ 
00030296012 & 54506.12279 & 0.59 & 2008-02-10 01:16:40 & 2008-02-10 04:36:58 & 873 & $1.8^{+0.5}_{-0.4}$ & $4.3^{+12.2}_{-1.6}$  	 \\ 
00030296013 & 54511.37989 & 0.03 & 2008-02-15 08:12:38 & 2008-02-15 10:01:25 & 1254 & $1.2^{+0.3}_{-0.2}$ &        --     	 \\ 
00030296014 & 54514.35488 & 0.28 & 2008-02-18 06:49:05 & 2008-02-18 10:12:57 & 1151 & $4.4^{+1.0}_{-0.8}$ & $22.6^{+240.6}_{-13.7}$  	 \\ 
00030296015 & 54516.02799 & 0.42 & 2008-02-20 00:33:38 & 2008-02-20 00:46:58 & 800 & $0.5^{+0.5}_{-0.2}$ & $1.1^{+5.0}_{-0.5}$  	 \\ 
00030296016 & 54519.94461 & 0.75 & 2008-02-23 21:44:31 & 2008-02-23 23:35:57 & 1667 & $0.9^{+0.3}_{-0.2}$ & $1.8^{+18.3}_{-0.7}$  	 \\ 
00030296017 & 54522.98995 & 0.01 & 2008-02-26 23:38:06 & 2008-02-26 23:52:56 & 890    &      --  &        --      	 \\ 
00030296018 & 54526.99264 & 0.35 & 2008-03-01 23:38:51 & 2008-03-01 23:59:59 & 1266  &      --  &        --     	 \\ % 
00030296019 & 54528.14534 & 0.44 & 2008-03-03 01:50:45 & 2008-03-03 05:07:56 & 1076  &      --  &        --     	 \\ % 
00030296020 & 54531.05549 & 0.69 & 2008-03-06 00:26:49 & 2008-03-06 02:12:58 & 1101  &      --  &        --     	 \\ 
00030296021 & 54535.44502 & 0.06 & 2008-03-10 09:05:42 & 2008-03-10 12:15:56 & 1261 & $27.2^{+3.6}_{-3.2}$ & $38.7^{+7.2}_{-4.4}$  	 \\ 
00030296022 & 54536.90404 & 0.18 & 2008-03-11 20:00:41 & 2008-03-11 23:22:56 & 2986 & $0.1^{+0.1}_{-0.0}$ & $1.7^{+20.1}_{-1.6}$  	 \\ 
00030296023 & 54537.88839 & 0.26 & 2008-03-12 18:46:44 & 2008-03-12 23:51:49 & 2776 & $5.6^{+0.7}_{-0.6}$ & $10.5^{+3.9}_{-1.7}$  	 \\ 
00030296024 & 54538.18389 & 0.29 & 2008-03-13 04:14:40 & 2008-03-13 04:34:56 & 1214 & $2.7^{+0.7}_{-0.5}$ & $6.0^{+10.6}_{-1.9}$  	 \\ 
00030296025 & 54540.19923 & 0.46 & 2008-03-15 04:37:49 & 2008-03-15 04:55:57 & 1088  &      --  &        --     	 \\ % 
00030296026 & 54541.33765 & 0.55 & 2008-03-16 07:56:54 & 2008-03-16 08:15:30 & 1116 & $4.8^{+0.9}_{-0.8}$ & $7.7^{+3.4}_{-1.3}$  	 \\ 
00030296027 & 54542.31110 & 0.64 & 2008-03-17 06:34:08 & 2008-03-17 08:21:48 & 1216 & $0.3^{+0.4}_{-0.2}$ & $0.5^{+2.9}_{-0.2}$  	 \\ 
00030296028 & 54543.47703 & 0.73 & 2008-03-18 06:31:54 & 2008-03-18 16:21:56 & 2342 & $0.3^{+0.2}_{-0.1}$ & $0.3^{+0.2}_{-0.1}$  	 \\ 
00030296029 & 54544.82184 & 0.85 & 2008-03-19 19:38:56 & 2008-03-19 19:47:57 & 542 & $0.8^{+0.6}_{-0.3}$ & $1.7^{+24.1}_{-0.8}$  	 \\ 
00306829000 & 54544.99873 & 0.86 & 2008-03-19 23:55:27 & 2008-03-20 00:00:53 & 326 & $8.7^{+2.9}_{-2.0}$ & $44.8^{+203.9}_{-27.1}$  	 \\ 
00030296030 & 54546.12308 & 0.96 & 2008-03-21 02:05:29 & 2008-03-21 03:48:58 & 835 & $1.6^{+1.5}_{-0.7}$ & $6.5^{+7.8}_{-4.5}$  	 \\ 
00030296031 & 54546.92654 & 0.02 & 2008-03-21 21:22:28 & 2008-03-21 23:05:57 & 835 &      --  &        --     		 \\ 
00030296032 & 54548.22937 & 0.13 & 2008-03-23 05:25:03 & 2008-03-23 05:35:30 & 627 & $0.3^{+0.4}_{-0.2}$ & $0.6^{+2.4}_{-0.3}$  	 \\ 
00030296033 & 54549.25965 & 0.22 & 2008-03-24 05:36:51 & 2008-03-24 06:50:56 & 1041 & $3.9^{+0.9}_{-0.7}$ & $5.5^{+1.9}_{-0.9}$  	 \\ 
00030296034 & 54566.73909 & 0.69 & 2008-04-10 15:17:37 & 2008-04-10 20:10:57 & 905 &      --  &        --     		 \\ 
00030296035 & 54588.93000 & 0.56 & 2008-05-02 22:12:26 & 2008-05-02 22:25:56 & 810 & $14.6^{+3.2}_{-2.6}$ & $21.4^{+9.2}_{-3.7}$  	 \\ 
00030296036 & 54592.86247 & 0.89 & 2008-05-06 20:40:14 & 2008-05-06 20:43:39 & 206 &      --  &        --     	  	 \\ 
00030296037 & 54596.41810 & 0.19 & 2008-05-10 09:53:10 & 2008-05-10 10:10:56 & 1066 & $3.0^{+0.7}_{-0.6}$ & $10.4^{+39.2}_{-4.9}$  	 \\ 
00030296038 & 54600.36325 & 0.52 & 2008-05-14 08:33:12 & 2008-05-14 08:52:58 & 1186  &      --  &        --     	 \\ % 
00030296039 & 54608.90984 & 0.24 & 2008-05-22 20:54:24 & 2008-05-22 22:45:56 & 2374 & $0.6^{+0.3}_{-0.2}$ & $1.1^{+5.2}_{-0.3}$  	 \\ 
00030296040 & 54612.88702 & 0.58 & 2008-05-26 21:05:40 & 2008-05-26 21:28:57 & 1397 & $2.8^{+0.6}_{-0.5}$ & $4.0^{+1.6}_{-0.7}$  	 \\ 
00030296041 & 54617.05938 & 0.93 & 2008-05-30 08:40:01 & 2008-05-31 18:10:58 & 995 & $2.0^{+1.4}_{-0.8}$ & $3.9^{+41.4}_{-1.5}$  	 \\ 
00030296042 & 54620.85094 & 0.25 & 2008-06-03 20:15:44 & 2008-06-03 20:34:57 & 1153 & $1.8^{+0.7}_{-0.5}$ & $2.0^{+0.7}_{-0.5}$  	 \\ 
00030296043 & 54624.67212 & 0.57 & 2008-06-07 16:05:45 & 2008-06-07 16:09:56 & 251 & $12.8^{+3.1}_{-2.5}$ & $24.0^{+25.8}_{-6.1}$  	 \\ 
00030296044 & 54628.38194 & 0.88 & 2008-06-11 08:22:02 & 2008-06-11 09:57:56 & 736  &      --  &        --     	 \\ % 
00030296046 & 54636.74855 & 0.58 & 2008-06-19 17:03:51 & 2008-06-19 18:51:58 & 1512  &      --  &        --     	 \\ % 
00030296047 & 54640.38837 & 0.89 & 2008-06-23 09:12:34 & 2008-06-23 09:25:56 & 802 & $1.6^{+0.9}_{-0.5}$ & $32.6^{+189.2}_{-28.9}$  	 \\ 
00030296048 & 54645.11642 & 0.29 & 2008-06-28 00:35:19 & 2008-06-28 04:59:57 & 970 & $2.4^{+0.6}_{-0.5}$ & $3.7^{+2.6}_{-0.8}$  	 \\ 
00030296049 & 54648.22275 & 0.55 & 2008-07-01 05:12:34 & 2008-07-01 05:28:57 & 983 & $2.5^{+0.7}_{-0.5}$ & $3.9^{+3.3}_{-0.9}$  	 \\ 
00030296050 & 54652.58477 & 0.92 & 2008-07-05 13:53:11 & 2008-07-05 14:10:57 & 1066 & $0.7^{+0.5}_{-0.3}$ & $0.7^{+0.5}_{-0.3}$  	 \\ 
00030296052 & 54660.44651 & 0.58 & 2008-07-13 09:51:01 & 2008-07-13 11:34:56 & 948 & $2.2^{+0.6}_{-0.5}$ & $5.3^{+13.4}_{-1.9}$  	 \\ 
00030296053 & 54664.56157 & 0.93 & 2008-07-17 13:26:34 & 2008-07-17 13:30:44 & 251  &      --  &        --     	 \\ % 
00030296055 & 54668.09847 & 0.22 & 2008-07-21 02:13:37 & 2008-07-21 02:29:57 & 980 & $2.2^{+0.7}_{-0.5}$ & $2.7^{+1.0}_{-0.5}$  	 \\ 
00030296056 & 54672.12657 & 0.56 & 2008-07-25 02:53:34 & 2008-07-25 03:10:57 & 1043 & $7.0^{+1.8}_{-1.4}$ &       --     	 \\ 	 
00030296057 & 54676.40988 & 0.92 & 2008-07-29 09:41:29 & 2008-07-29 09:58:57 & 1020  &      --  &        --     	 \\ % 
00030296058 & 54680.57905 & 0.27 & 2008-08-02 13:00:42 & 2008-08-02 14:46:58 & 933 & $7.9^{+1.6}_{-1.3}$ & $15.0^{+11.7}_{-3.3}$  	 \\ 
00030296059 & 54684.09551 & 0.57 & 2008-08-06 02:09:08 & 2008-08-06 02:25:56 & 1008 & $28.0^{+4.5}_{-3.9}$ & $40.9^{+9.4}_{-5.3}$  	 \\ 
00030296060 & 54688.04478 & 0.90 & 2008-08-10 00:57:00 & 2008-08-10 01:11:57 & 898 & $4.0^{+1.6}_{-1.1}$ & $5.5^{+5.5}_{-1.3}$  	 \\ 
00030296061 & 54692.80424 & 0.30 & 2008-08-14 19:10:16 & 2008-08-14 19:25:56 & 940  &      --  &        --     	 \\ % 
00030296062 & 54696.48327 & 0.61 & 2008-08-18 11:27:53 & 2008-08-18 11:43:56 & 963  &      --  &        --     	 \\ % 
00030296063 & 54700.62840 & 0.96 & 2008-08-22 07:01:50 & 2008-08-22 23:07:57 & 755  &      --  &        --     	 \\ % 
\noalign{\smallskip} % 
 \hline
  \end{tabular}
  \end{center}
  \end{table*}

\setcounter{table}{2}		
 \begin{table*} 	
\tabcolsep 4pt         
   \begin{center}
  \caption{Observation log for IGR~J16479$-$4514. Continued.} 
 \begin{tabular}{ l l l ll r rr   } 
 \hline 
 \hline 
 \noalign{\smallskip} 

 ObsID           &  MJD   & Phase & Start time  (UT)                     & End time   (UT)                 & Exposure  & Flux$^a$ & Flux$^b$  \\ 
                      &           &          &  (yyyy-mm-dd hh:mm:ss)  & (yyyy-mm-dd hh:mm:ss)  &(s)              & (erg\,cm$^{-2}$\,s$^{-1}$) & (erg\,cm$^{-2}$\,s$^{-1}$)               \\
  \noalign{\smallskip} 
 \hline 
 \noalign{\smallskip} 
00030296065 & 54708.09900 & 0.59 & 2008-08-30 01:29:10 & 2008-08-30 03:15:56 & 1236 & $5.5^{+1.2}_{-1.0}$ & $7.0^{+1.2}_{-1.0}$  	 \\ 
00030296066 & 54720.97606 & 0.67 & 2008-09-11 23:18:06 & 2008-09-11 23:32:56 & 890 & $1.0^{+0.5}_{-0.3}$ & $8.7^{+84.4}_{-6.9}$  	 \\ 
00030296067 & 54724.32964 & 0.96 & 2008-09-15 07:46:25 & 2008-09-15 08:02:56 & 990 & $0.7^{+0.3}_{-0.2}$ & $3.6^{+37.8}_{-2.4}$  	 \\ 
00030296070 & 54736.24306 & 0.96 & 2008-09-27 05:41:07 & 2008-09-27 05:58:55 & 1066  &      --  &        --     	 \\ % 
00030296071 & 54740.66567 & 0.33 & 2008-10-01 15:49:11 & 2008-10-01 16:07:57 & 1126 & $0.4^{+0.5}_{-0.2}$ & $16.0^{+173.0}_{-15.4}$  	 \\ 
00030296073 & 54751.97333 & 0.28 & 2008-10-12 23:14:15 & 2008-10-12 23:28:55 & 880 & $1.2^{+0.5}_{-0.4}$ & $4.4^{+13.6}_{-2.5}$  	 \\ 
00030296074 & 54756.09479 & 0.63 & 2008-10-17 01:27:07 & 2008-10-17 03:05:57 & 466  &      --  &        --     	 \\ % 
00030296075 & 54760.20390 & 0.98 & 2008-10-21 04:46:15 & 2008-10-21 05:00:57 & 883 & $1.8^{+0.7}_{-0.5}$ & $2.4^{+1.8}_{-0.5}$  	 \\ 
00030296076 & 54764.95512 & 0.38 & 2008-10-25 22:48:38 & 2008-10-25 23:02:05 & 807 & $9.4^{+1.9}_{-1.5}$ & $15.7^{+8.9}_{-3.1}$  	 \\ 
00341452000 & 54860.32175 & 0.40 & 2009-01-29 06:47:34 & 2009-01-29 08:39:03 & 2660 & $19.2^{+2.3}_{-2.0}$ & $27.7^{+4.6}_{-2.9}$  	 \\ 
00030296077 & 54860.80421 & 0.44 & 2009-01-29 15:59:43 & 2009-01-29 22:36:23 & 5927 & $3.9^{+0.4}_{-0.3}$ & $6.2^{+1.1}_{-0.7}$  	 \\ 
00030296078 & 54861.25266 & 0.48 & 2009-01-30 05:05:44 & 2009-01-30 07:01:55 & 2482 & $4.0^{+0.6}_{-0.5}$ & $6.7^{+2.3}_{-1.0}$  	 \\ 
00030296079 & 54862.05578 & 0.55 & 2009-01-31 00:23:44 & 2009-01-31 02:17:57 & 1366 & $0.6^{+0.3}_{-0.2}$ & $1.0^{+4.5}_{-0.3}$  	 \\ 
00030296080 & 54863.86170 & 0.70 & 2009-02-01 19:50:45 & 2009-02-01 21:30:56 & 2043 & $0.8^{+0.3}_{-0.2}$ & $0.9^{+0.3}_{-0.2}$  	 \\ 
00030296081 & 54864.42608 & 0.75 & 2009-02-02 08:46:11 & 2009-02-02 11:40:56 & 1903 & $0.4^{+0.3}_{-0.1}$ & $1.1^{+6.2}_{-0.6}$  	 \\ 
00030296082 & 54865.83880 & 0.87 & 2009-02-03 18:25:47 & 2009-02-03 21:49:57 & 1236  &      --  &        --     	 \\ % 
00030296083 & 54866.10478 & 0.89 & 2009-02-04 00:51:49 & 2009-02-04 04:09:55 & 1409 &      --  &        --     	 	 \\ 
00030296084 & 54867.10934 & 0.98 & 2009-02-05 02:24:24 & 2009-02-05 02:50:29 & 1507 & $2.0^{+0.6}_{-0.5}$ & $3.1^{+3.6}_{-0.7}$  	 \\ 
00030296085 & 54868.11879 & 0.06 & 2009-02-06 01:13:11 & 2009-02-06 04:28:56 & 486 & $2.3^{+1.3}_{-0.8}$ & $2.7^{+1.1}_{-0.8}$  	 \\ 
00030296087 & 54870.85766 & 0.29 & 2009-02-08 18:54:07 & 2009-02-08 22:15:56 & 1620 & $22.7^{+2.6}_{-2.3}$ & $32.9^{+4.9}_{-3.2}$  	 \\ 
00030296088 & 54872.23515 & 0.41 & 2009-02-10 03:10:17 & 2009-02-10 08:06:56 & 1565  &      --  &        --     	 \\ % 
00030296089 & 54873.80090 & 0.54 & 2009-02-11 15:47:38 & 2009-02-11 22:38:56 & 2247 & $7.9^{+1.1}_{-0.9}$ & $17.5^{+12.1}_{-4.1}$  	 \\ 
00030296090 & 54874.97612 & 0.64 & 2009-02-12 22:30:17 & 2009-02-13 00:20:55 & 1695  &      --  &        --     	 \\ % 
00030296091 & 54875.24659 & 0.66 & 2009-02-13 05:04:17 & 2009-02-13 06:45:53 & 1314  &      --  &        --     	 \\ % 
00030296092 & 54876.31004 & 0.75 & 2009-02-14 06:34:00 & 2009-02-14 08:18:55 & 1271 & $2.7^{+0.8}_{-0.6}$ & $3.0^{+0.7}_{-0.5}$  	 \\ 
00030296093 & 54877.14570 & 0.82 & 2009-02-15 03:19:40 & 2009-02-15 03:39:56 & 1216 &      --  &        -- 	 \\ 
00030296094 & 54878.38930 & 0.93 & 2009-02-16 08:26:28 & 2009-02-16 10:14:41 & 1126 & $2.5^{+0.7}_{-0.5}$ & $6.2^{+21.6}_{-2.4}$  	 \\ 
00030296095 & 54879.02144 & 0.98 & 2009-02-17 00:22:54 & 2009-02-17 00:38:49 & 955  &      --  &        --     	 \\ % 
00030296096 & 54881.92902 & 0.22 & 2009-02-19 21:23:38 & 2009-02-19 23:11:55 & 1201 & $4.6^{+1.0}_{-0.8}$ & $7.1^{+3.6}_{-1.3}$  	 \\ 
00030296097 & 54885.67519 & 0.54 & 2009-02-23 15:15:36 & 2009-02-23 17:08:57 & 1933  &      --  &        --     	 \\ % 
00030296098 & 54888.89594 & 0.81 & 2009-02-26 20:38:21 & 2009-02-26 22:21:57 & 910  &      --  &        --     	 \\ % 
00030296099 & 54891.10343 & 1.00 & 2009-03-01 01:36:57 & 2009-03-01 03:20:56 & 1018 & $0.5^{+0.3}_{-0.2}$ & $1.5^{+16.7}_{-0.9}$  	 \\ 
00030296101 & 54898.08683 & 0.58 & 2009-03-08 01:58:10 & 2009-03-08 02:11:57 & 825 & $2.0^{+0.6}_{-0.5}$ & $3.1^{+3.3}_{-0.7}$  	 \\ 
00030296102 & 54902.50499 & 0.95 & 2009-03-12 12:00:25 & 2009-03-12 12:13:55 & 810  &      --  &        --     	 \\ % 
00030296103 & 54905.73228 & 0.23 & 2009-03-15 17:26:02 & 2009-03-15 17:42:55 & 1013  &      --  &        --     	 \\ % 
00030296104 & 54909.95111 & 0.58 & 2009-03-19 22:41:21 & 2009-03-19 22:57:49 & 988  &      --  &        --     	 \\ % 
00030296105 & 54913.22458 & 0.86 & 2009-03-23 05:11:50 & 2009-03-23 05:34:56 & 1361  &      --  &        --     	 \\ % 
00030296106 & 54916.97633 & 0.17 & 2009-03-26 23:17:54 & 2009-03-26 23:33:55 & 950 & $1.9^{+0.5}_{-0.4}$ & $3.6^{+4.7}_{-1.0}$  	 \\ 
00030296107 & 54919.88405 & 0.42 & 2009-03-29 20:20:09 & 2009-03-29 22:05:54 & 1171 & $1.6^{+0.5}_{-0.4}$ & $2.7^{+4.6}_{-0.8}$  	 \\ 
00030296108 & 54923.49186 & 0.72 & 2009-04-02 10:51:37 & 2009-04-02 12:44:56 & 1221 & $1.2^{+0.4}_{-0.3}$ & $2.7^{+6.5}_{-1.0}$  	 \\ 
00030296110 & 54930.96239 & 0.35 & 2009-04-09 22:57:45 & 2009-04-09 23:13:55 & 970 & $0.6^{+0.6}_{-0.3}$ & $0.9^{+5.2}_{-0.3}$  	 \\ 
00030296111 & 54933.03779 & 0.53 & 2009-04-12 00:46:55 & 2009-04-12 01:01:53 & 870 & $2.2^{+0.6}_{-0.5}$ & $4.2^{+5.9}_{-1.1}$  	 \\ 
00030296112 & 54937.08095 & 0.87 & 2009-04-16 01:03:11 & 2009-04-16 02:49:57 & 875  &      --  &        --     	 \\ % 
00030296114 & 54944.31288 & 0.47 & 2009-04-23 06:35:09 & 2009-04-23 08:25:56 & 1151 & $0.6^{+0.4}_{-0.2}$ & $1.0^{+5.3}_{-0.3}$  	 \\ 
00030296116 & 54954.85973 & 0.36 & 2009-05-03 20:31:04 & 2009-05-03 20:44:57 & 832 & $3.4^{+1.0}_{-0.8}$ & $5.2^{+5.6}_{-1.2}$  	 \\ 
00030296117 & 54958.32941 & 0.65 & 2009-05-07 07:45:45 & 2009-05-07 08:02:55 & 1031 & $0.8^{+0.7}_{-0.4}$ & $0.9^{+0.7}_{-0.4}$  	 \\ 
00030296118 & 54965.93235 & 0.29 & 2009-05-14 21:31:14 & 2009-05-14 23:13:55 & 782  &      --  &        --     	 \\ % 
00030296119 & 54968.76808 & 0.53 & 2009-05-17 18:25:08 & 2009-05-17 18:26:56 & 108  &      --  &        --     	 \\ % 
00030296120 & 54972.14231 & 0.82 & 2009-05-21 02:33:55 & 2009-05-21 04:15:56 & 837  &      --  &        --     	 \\ % 
00030296121 & 54979.70379 & 0.45 & 2009-05-28 16:01:58 & 2009-05-28 17:44:56 & 832 & $6.1^{+1.7}_{-1.3}$ & $9.6^{+7.8}_{-1.9}$  	 \\ 
00030296122 & 54986.40167 & 0.02 & 2009-06-04 09:00:39 & 2009-06-04 10:22:55 & 70   &      --  &        --     	 \\ 
00030296123 & 54992.76784 & 0.55 & 2009-06-10 14:19:26 & 2009-06-10 22:31:56 & 1086 & $1.2^{+0.6}_{-0.4}$ & $1.3^{+0.6}_{-0.4}$  	 \\ 
00030296124 & 54993.42420 & 0.61 & 2009-06-11 09:17:45 & 2009-06-11 11:03:56 & 1176 & $0.5^{+0.4}_{-0.2}$ & $0.7^{+0.4}_{-0.2}$  	 \\ 
00030296125 & 54996.61213 & 0.88 & 2009-06-14 12:56:59 & 2009-06-14 16:25:56 & 1354 & $2.1^{+0.8}_{-0.6}$ & $2.5^{+1.1}_{-0.6}$  	 \\ 
00030296127 & 55004.93965 & 0.58 & 2009-06-22 21:41:14 & 2009-06-22 23:24:56 & 915  &      --  &        --     	 \\ % 
00030296128 & 55007.65543 & 0.81 & 2009-06-25 15:35:41 & 2009-06-25 15:51:57 & 975 & $1.5^{+0.5}_{-0.4}$ & $4.7^{+29.0}_{-2.3}$  	 \\ 
\noalign{\smallskip} % 
 \hline
  \end{tabular}
  \end{center}
  \end{table*}

\setcounter{table}{2}		
 \begin{table*} 	
\tabcolsep 4pt         
   \begin{center}
  \caption{Observation log for IGR~J16479$-$4514. Continued.} 
 \begin{tabular}{ l l l ll r rr   } 
 \hline 
 \hline 
 \noalign{\smallskip} 

 ObsID           &  MJD   & Phase & Start time  (UT)                     & End time   (UT)                 & Exposure  & Flux$^a$ & Flux$^b$  \\ 
                      &           &          &  (yyyy-mm-dd hh:mm:ss)  & (yyyy-mm-dd hh:mm:ss)  &(s)              & (erg\,cm$^{-2}$\,s$^{-1}$) & (erg\,cm$^{-2}$\,s$^{-1}$)               \\
  \noalign{\smallskip} 
 \hline 
 \noalign{\smallskip} 
00030296129 & 55011.38840 & 0.12 & 2009-06-29 02:03:37 & 2009-06-29 16:34:56 & 419  &     --  &        --      	 \\ 
 00030296130 & 55014.96280 & 0.42 & 2009-07-02 22:56:55 & 2009-07-02 23:15:56 & 1141  &      --  &        --     	 \\ % 
00030296131 & 55018.97409 & 0.76 & 2009-07-06 23:03:26 & 2009-07-06 23:41:56 & 943  &      --  &        --     	 \\ % 
00030296132 & 55021.92975 & 0.01 & 2009-07-09 20:41:43 & 2009-07-09 23:55:56 & 394  &      --  &        --     	 \\ % 
00030296133 & 55024.35952 & 0.21 & 2009-07-12 07:45:28 & 2009-07-12 09:29:57 & 1221 & $1.6^{+0.5}_{-0.3}$ & $5.9^{+34.1}_{-3.1}$  	 \\ 
00030296134 & 55031.75641 & 0.83 & 2009-07-19 18:04:31 & 2009-07-19 18:13:55 & 564  &      --  &        --     	 \\ % 
00030296135 & 55035.67416 & 0.16 & 2009-07-23 15:17:38 & 2009-07-23 17:03:55 & 702 & $4.4^{+1.3}_{-1.0}$ & $7.4^{+10.8}_{-1.9}$  	 \\ 
00030296136 & 55038.63799 & 0.41 & 2009-07-26 13:36:27 & 2009-07-26 17:00:56 & 973  &      --  &        --     	 \\ % 
00030296137 & 55042.11395 & 0.71 & 2009-07-30 01:03:15 & 2009-07-30 04:24:55 & 1622 & $5.5^{+1.2}_{-1.0}$ & $7.1^{+2.5}_{-1.2}$  	 \\ 
00030296138 & 55045.81150 & 0.02 & 2009-08-02 19:20:11 & 2009-08-02 19:36:56 & 1005 & $2.1^{+0.6}_{-0.5}$ & $3.5^{+3.7}_{-0.8}$  	 \\ 
00030296139 & 55049.57839 & 0.33 & 2009-08-06 12:58:48 & 2009-08-06 14:46:56 & 1178 & $2.3^{+0.5}_{-0.4}$ & $4.2^{+4.2}_{-1.0}$  	 \\ 
00030296141 & 55056.11021 & 0.88 & 2009-08-13 02:28:40 & 2009-08-13 02:48:44 & 817 & $5.5^{+1.3}_{-1.1}$ & $8.1^{+4.6}_{-1.5}$  	 \\ 
00030296142 & 55063.74699 & 0.53 & 2009-08-20 17:46:22 & 2009-08-20 18:04:56 & 1111 & $0.7^{+0.4}_{-0.2}$ & $1.3^{+6.3}_{-0.5}$  	 \\ 
00030296143 & 55066.49957 & 0.76 & 2009-08-23 11:50:58 & 2009-08-23 12:07:56 & 1008  &      --  &        --     	 \\ % 
00030296144 & 55070.24617 & 0.07 & 2009-08-27 05:46:00 & 2009-08-27 06:02:56 & 1015 & $1.3^{+0.6}_{-0.4}$ & $1.8^{+8.0}_{-0.5}$  	 \\ 
00030296145 & 55073.25458 & 0.33 & 2009-08-30 05:52:16 & 2009-08-30 06:20:56 & 1720 & $0.4^{+0.2}_{-0.1}$ & $1.5^{+25.7}_{-0.9}$  	 \\ 
00030296146 & 55084.63249 & 0.29 & 2009-09-10 15:02:37 & 2009-09-10 15:18:57 & 980 & $0.1^{+0.3}_{-0.1}$ & $0.2^{+0.7}_{-0.2}$  	 \\
00030296147 & 55087.64497 & 0.54 & 2009-09-13 15:20:33 & 2009-09-13 15:36:56 & 983 & $0.4^{+0.3}_{-0.2}$ & $1.3^{+8.4}_{-0.8}$  	 \\ 
00030296148 & 55091.12292 & 0.83 & 2009-09-17 02:48:03 & 2009-09-17 03:05:56 & 1073  &      --  &        --     	 \\ % 
00030296149 & 55094.28585 & 0.10 & 2009-09-20 06:00:18 & 2009-09-20 07:42:55 & 802 & $23.6^{+4.7}_{-3.9}$ & $40.8^{+38.1}_{-9.0}$  	 \\ 
00030296150 & 55101.65828 & 0.72 & 2009-09-27 14:53:55 & 2009-09-27 16:41:56 & 1128  &      --  &        --     	 \\ % 
00030296151 & 55105.91559 & 0.08 & 2009-10-01 21:50:57 & 2009-10-01 22:05:57 & 900 & $1.4^{+0.6}_{-0.4}$ & $2.6^{+9.2}_{-1.0}$  	 \\ 
00030296152 & 55108.11932 & 0.26 & 2009-10-04 02:43:41 & 2009-10-04 02:59:56 & 963  &      --  &        --     	 \\ % 
00030296153 & 55112.13932 & 0.60 & 2009-10-08 03:12:18 & 2009-10-08 03:28:56 & 998 & $0.9^{+0.6}_{-0.3}$ & $2.5^{+18.4}_{-1.4}$  	 \\ 
00030296154 & 55122.57440 & 0.48 & 2009-10-18 13:39:20 & 2009-10-18 13:54:55 & 905 & $1.4^{+0.5}_{-0.3}$ & $9.0^{+79.6}_{-6.4}$  	 \\ 
00030296155 & 55129.33971 & 0.05 & 2009-10-25 08:04:26 & 2009-10-25 08:13:55 & 569 & $16.6^{+3.7}_{-3.0}$ & $33.1^{+41.3}_{-8.8}$  	 \\ 
\noalign{\smallskip}
 \hline
  \end{tabular}
  \end{center}
  \end{table*}

              %%%%%%%%%%%%%%%%%%%%%%%%%%%%%%%%%%%%%%%%%%%%%%%%%%%%%%%%%% TABLE A4
              %\input{table_a4_16493_xrt_log_phase.tex}  %  16493_xrt_log 
              %%%%%%%%%%%%%%%%%%%%%%%%%%%%%%%%%%%%%%%%%%%%%%%%%%%%%%%%%% 

 \setcounter{table}{3}
 \begin{table*}
\tabcolsep 4pt
   \begin{center}
  \caption{Same as Table~\ref{supero:tab:swift_xrt_log_16479}  for IGR~J16493$-$4348.
 \label{supero:tab:swift_xrt_log_16493}}
  \begin{tabular}{ l l l ll r rr  }
 \hline
 ObsID           &  MJD   & Phase & Start time  (UT)                     & End time   (UT)                 & Exposure  & Flux & Flux      \\
                      &           &          &  (yyyy-mm-dd hh:mm:ss)  & (yyyy-mm-dd hh:mm:ss)  &(s)              & (erg\,cm$^{-2}$\,s$^{-1}$) & (erg\,cm$^{-2}$\,s$^{-1}$)                \\
% \noalign{\smallskip}
 \hline
\noalign{\smallskip}
%
% 
%
% Romano, 2015, JHEAP, 7, 126;   Kretschmar et al., 2019, NewAR, 86, 101546 
00030379003 & 56676.84697 & 0.17 & 2014-01-19 20:11:21 & 2014-01-19 20:27:54 & 993 & $2.9^{+0.7}_{-0.5}$ & $6.2^{+10.3}_{-1.9}$  	 \\ 
00030379004 & 56679.12236 & 0.28 & 2014-01-22 02:49:29 & 2014-01-22 03:02:54 & 805 & $1.8^{+0.9}_{-0.6}$ & $13.3^{+78.5}_{-10.0}$  	 \\ 
00030379005 & 56683.05763 & 0.48 & 2014-01-26 01:10:04 & 2014-01-26 01:35:54 & 1550 & $3.9^{+0.7}_{-0.6}$ & $5.6^{+1.6}_{-0.8}$  	 \\ 
00030379006 & 56686.25279 & 0.64 & 2014-01-29 05:56:05 & 2014-01-29 06:11:56 & 950 &      --  &        --  	 \\ 
00030379007 & 56690.06221 & 0.83 & 2014-02-02 01:21:14 & 2014-02-02 01:37:55 & 1000 & $4.8^{+1.1}_{-0.9}$ & $5.8^{+1.0}_{-0.9}$  	 \\ 
00030379008 & 56693.06911 & 0.97 & 2014-02-05 01:31:07 & 2014-02-05 01:47:55 & 1008 & $2.4^{+0.6}_{-0.5}$ & $8.6^{+45.2}_{-4.3}$  	 \\ 
00030379009 & 56697.26632 & 0.18 & 2014-02-09 06:14:06 & 2014-02-09 06:32:54 & 1128 & $2.9^{+0.7}_{-0.6}$ & $3.5^{+0.6}_{-0.6}$  	 \\ 
00030379010 & 56704.39711 & 0.54 & 2014-02-16 01:12:45 & 2014-02-16 17:50:56 & 1221 & $2.1^{+0.7}_{-0.5}$ & $4.5^{+19.5}_{-1.8}$  	 \\ 
00030379011 & 56707.20263 & 0.68 & 2014-02-19 04:43:41 & 2014-02-19 04:59:53 & 973 & $3.3^{+0.9}_{-0.7}$ & $6.0^{+9.4}_{-1.6}$  	 \\ 
00030379012 & 56711.75433 & 0.91 & 2014-02-23 12:26:32 & 2014-02-23 23:45:55 & 865 & $0.5^{+0.5}_{-0.3}$ & $0.9^{+8.8}_{-0.4}$  	 \\ 
00030379013 & 56714.47428 & 0.04 & 2014-02-26 09:44:00 & 2014-02-26 13:01:55 & 980 & $3.9^{+1.3}_{-1.0}$ & $5.1^{+2.6}_{-1.0}$  	 \\ 
00030379014 & 56718.86194 & 0.26 & 2014-03-02 20:33:27 & 2014-03-02 20:48:55 & 925  &      --  &        --     	 \\ % 
00030379016 & 56725.32999 & 0.58 & 2014-03-09 07:47:26 & 2014-03-09 08:02:54 & 928  &      --  &        --     	 \\ % 
00030379017 & 56728.53286 & 0.74 & 2014-03-12 12:38:43 & 2014-03-12 12:55:54 & 1031 & $2.3^{+0.7}_{-0.5}$ & $3.4^{+2.4}_{-0.7}$  	 \\ 
00030379018 & 56732.14626 & 0.92 & 2014-03-16 03:22:20 & 2014-03-16 03:38:53 & 993 & $5.5^{+1.5}_{-1.3}$ & $8.2^{+5.5}_{-1.9}$  	 \\ 
00030379019 & 56735.61277 & 0.09 & 2014-03-19 14:33:50 & 2014-03-19 14:50:55 & 1025 & $5.4^{+1.2}_{-1.0}$ & $10.4^{+14.8}_{-2.9}$  	 \\ 
00030379020 & 56739.61033 & 0.29 & 2014-03-23 14:35:25 & 2014-03-23 14:42:19 & 414  &      --  &        --     	 \\ % 
00030379021 & 56742.07266 & 0.42 & 2014-03-26 01:34:19 & 2014-03-26 01:54:55 & 1236 & $1.9^{+0.5}_{-0.4}$ & $2.8^{+2.0}_{-0.6}$  	 \\ 
00030379022 & 56746.78347 & 0.65 & 2014-03-30 17:58:28 & 2014-03-30 19:37:54 & 953  &      --  &        --     	 \\ % 
00030379023 & 56749.88149 & 0.81 & 2014-04-02 21:00:45 & 2014-04-02 21:17:55 & 1031 & $5.3^{+1.1}_{-0.9}$ & $6.9^{+1.5}_{-1.0}$  	 \\ 
00030379024 & 56753.13446 & 0.97 & 2014-04-06 00:03:17 & 2014-04-06 06:23:56 & 875  &      --  &        --     	 \\ % 
00030379025 & 56756.73465 & 0.15 & 2014-04-09 17:29:51 & 2014-04-09 17:45:54 & 963 & $9.9^{+1.8}_{-1.5}$ & $13.0^{+2.4}_{-1.7}$  	 \\ 
00030379026 & 56760.79934 & 0.35 & 2014-04-13 19:03:09 & 2014-04-13 19:18:57 & 945 & $3.1^{+1.0}_{-0.7}$ & $6.9^{+24.6}_{-2.4}$  	 \\ 
00030379027 & 56763.54492 & 0.49 & 2014-04-16 12:56:25 & 2014-04-16 13:12:55 & 990 & $5.8^{+1.6}_{-1.2}$ & $8.3^{+6.6}_{-1.6}$  	 \\ 
00030379028 & 56767.76756 & 0.70 & 2014-04-20 17:31:39 & 2014-04-20 19:18:54 & 248  &      --  &        --     	 \\ % 
00030379029 & 56770.53186 & 0.83 & 2014-04-23 12:37:51 & 2014-04-23 12:53:54 & 963 & $3.3^{+1.4}_{-1.0}$ & $3.6^{+1.0}_{-0.8}$  	 \\ 
00030379030 & 56774.74429 & 0.04 & 2014-04-27 17:43:37 & 2014-04-27 17:59:55 & 978 & $2.9^{+0.8}_{-0.6}$ & $6.8^{+7.2}_{-2.5}$  	 \\ 
00030379031 & 56777.21166 & 0.17 & 2014-04-30 03:27:41 & 2014-04-30 06:41:54 & 1013 & $3.1^{+0.8}_{-0.6}$ & $4.2^{+1.4}_{-0.7}$  	 \\ 
00030379032 & 56781.14707 & 0.36 & 2014-05-04 03:22:39 & 2014-05-04 03:40:54 & 1096 & $1.8^{+0.6}_{-0.5}$ & $2.7^{+2.7}_{-0.6}$  	 \\
00030379033 & 56784.74352 & 0.54 & 2014-05-07 17:42:24 & 2014-05-07 17:58:55 & 990 & $2.0^{+0.6}_{-0.4}$ & $4.5^{+11.1}_{-1.6}$  	 \\ 
00030379034 & 56788.47134 & 0.73 & 2014-05-11 11:10:33 & 2014-05-11 11:26:54 & 980 & $4.7^{+1.1}_{-0.9}$ & $6.5^{+2.3}_{-1.1}$  	 \\ 
00030379035 & 56791.06951 & 0.86 & 2014-05-14 01:32:15 & 2014-05-14 01:47:55 & 940 & $5.4^{+1.3}_{-1.0}$ & $7.6^{+3.2}_{-1.3}$  	 \\ 
00030379039 & 56805.67506 & 0.59 & 2014-05-28 16:03:15 & 2014-05-28 16:20:55 & 1061 & $1.1^{+0.6}_{-0.4}$ & $2.2^{+12.3}_{-0.9}$  	 \\ 
00030379040 & 56809.80156 & 0.79 & 2014-06-01 19:05:34 & 2014-06-01 19:22:55 & 1041 & $1.2^{+0.4}_{-0.3}$ & $23.4^{+206.4}_{-20.3}$  	 \\ 
00030379044 & 56823.45513 & 0.47 & 2014-06-15 10:47:50 & 2014-06-15 11:02:55 & 905 & $1.2^{+0.6}_{-0.4}$ & $1.8^{+4.2}_{-0.5}$  	 \\ 
00030379045 & 56826.32881 & 0.61 & 2014-06-18 07:45:13 & 2014-06-18 08:01:56 & 993 & $2.8^{+1.1}_{-0.8}$ & $7.6^{+36.3}_{-3.6}$  	 \\ 
00030379046 & 56830.89013 & 0.84 & 2014-06-22 04:22:40 & 2014-06-23 14:20:54 & 1186 & $2.3^{+0.7}_{-0.5}$ & $2.7^{+0.6}_{-0.5}$  	 \\ 
00030379047 & 56833.06993 & 0.95 & 2014-06-25 01:34:30 & 2014-06-25 01:46:54 & 745 & $4.4^{+1.1}_{-0.9}$ & $6.8^{+4.7}_{-1.4}$  	 \\ 
00030379048 & 56837.85812 & 0.19 & 2014-06-29 20:27:28 & 2014-06-29 20:43:54 & 985 & $6.6^{+1.2}_{-1.0}$ & $9.5^{+2.6}_{-1.3}$  	 \\ 
00030379049 & 56840.40633 & 0.32 & 2014-07-02 00:03:17 & 2014-07-02 19:26:55 & 1068 & $0.7^{+0.5}_{-0.3}$ & $1.1^{+8.9}_{-0.5}$  	 \\ 
00030379050 & 56844.53560 & 0.52 & 2014-07-06 12:45:35 & 2014-07-06 12:56:55 & 679 & $3.9^{+1.6}_{-1.1}$ & $4.4^{+1.4}_{-1.0}$  	 \\ 
00030379051 & 56847.38144 & 0.66 & 2014-07-09 09:08:36 & 2014-07-09 09:09:56 & 80  &      --  &        --     	 \\ % 
00030379052 & 56851.25188 & 0.86 & 2014-07-13 05:54:29 & 2014-07-13 06:10:54 & 985 & $3.8^{+0.8}_{-0.7}$ & $6.2^{+4.2}_{-1.3}$  	 \\ 
00030379053 & 56854.75888 & 0.03 & 2014-07-16 15:41:39 & 2014-07-16 20:43:55 & 750  &      --  &        --     	 \\ % 
00030379054 & 56858.59527 & 0.22 & 2014-07-20 14:08:27 & 2014-07-20 14:25:55 & 1048 & $1.6^{+0.6}_{-0.4}$ & $2.4^{+2.3}_{-0.5}$  	 \\ 
00030379055 & 56861.72998 & 0.38 & 2014-07-23 17:27:26 & 2014-07-23 17:34:54 & 449  &      --  &        --     	 \\ % 
00030379056 & 56865.17643 & 0.55 & 2014-07-27 04:05:12 & 2014-07-27 04:22:55 & 1063 & $3.0^{+0.8}_{-0.7}$ & $3.4^{+0.7}_{-0.6}$  	 \\ 
00030379057 & 56868.18760 & 0.70 & 2014-07-30 04:21:23 & 2014-07-30 04:38:53 & 1051  &      --  &        --     	 \\ % 
00030379058 & 56872.85217 & 0.93 & 2014-08-03 20:18:24 & 2014-08-03 20:35:55 & 1046 & $6.6^{+1.6}_{-1.2}$ & $12.4^{+15.2}_{-3.2}$  	 \\ 
00030379059 & 56875.57199 & 0.07 & 2014-08-06 13:35:23 & 2014-08-06 13:51:56 & 993  &      --  &        --     	 \\ % 
00030379060 & 56879.78244 & 0.28 & 2014-08-10 18:38:31 & 2014-08-10 18:54:54 & 983 & $1.1^{+0.5}_{-0.3}$ & $2.9^{+22.5}_{-1.3}$  	 \\ 
00030379061 & 56882.03947 & 0.39 & 2014-08-13 00:48:46 & 2014-08-13 01:04:54 & 968  &      --  &        --     	 \\ % 
00030379062 & 56886.64423 & 0.62 & 2014-08-17 15:24:31 & 2014-08-17 15:30:55 & 379  &      --  &        --     	 \\ 
00030379063 & 56889.70647 & 0.77 & 2014-08-20 16:49:41 & 2014-08-20 17:04:56 & 915 & $6.8^{+1.7}_{-1.4}$ & $9.1^{+2.9}_{-1.5}$  	 \\ 
00030379065 & 56896.17544 & 0.09 & 2014-08-27 04:04:38 & 2014-08-27 04:20:54 & 958  &      --  &        --  	 \\ 
00030379066 & 56900.31039 & 0.30 & 2014-08-31 07:19:01 & 2014-08-31 07:34:53 & 953 & $1.7^{+0.5}_{-0.4}$ & $4.8^{+25.1}_{-2.1}$  	 \\ 
00030379067 & 56903.50922 & 0.46 & 2014-09-03 12:04:37 & 2014-09-03 12:21:55 & 1038 & $2.9^{+0.7}_{-0.6}$ & $4.7^{+4.3}_{-1.0}$  	 \\ 

\noalign{\smallskip}
 \hline
  \end{tabular}

  \end{center}
  \end{table*}

   %%%%%%%%%%%%%%%%%%%%%%%%%%%%%%%%%%%%%%%%%%%%%%%%%%%%%%%%%
    \section{Additional Material} \label{supero:appendix2} 
   %%%%%%%%%%%%%%%%%%%%%%%%%%%%%%%%%%%%%%%%%%%%%%%%%%%%%%%%%

As reported in Sect.~\ref{supero:2s}, the X-ray light curves of \mysrc{}
show variability on different timescales, namely, 
the spin period  P$_{\rm spin}\sim9800$\,s,   
the orbital period P$_{\rm orb}$=11.5983\,d, and 
the superorbital period P$_{\rm sup}$=30.76\,d. 
In order to clarify the effects of binning of the data on these different timescales
on our results we also calculated the spin phase for each observation (Table~\ref{supero:tab:swift_xrt_log_s0114}).
We note that the typical exposure time of our observations is generally a factor of 10 
shorter than  P$_{\rm spin}$, so we addressed the possibility that 
in each orbital phase bin the spin phases would not be equally represented.

\begin{figure*}%%%%%%%%%%%%%%%%%%%%%%%%%%%%%%%%%%%%%%%%%%%%%%%%%%%%%%%%% FIGURE B1

 \hspace{-0.5truecm}
  \includegraphics[width=18.5cm,angle=0]{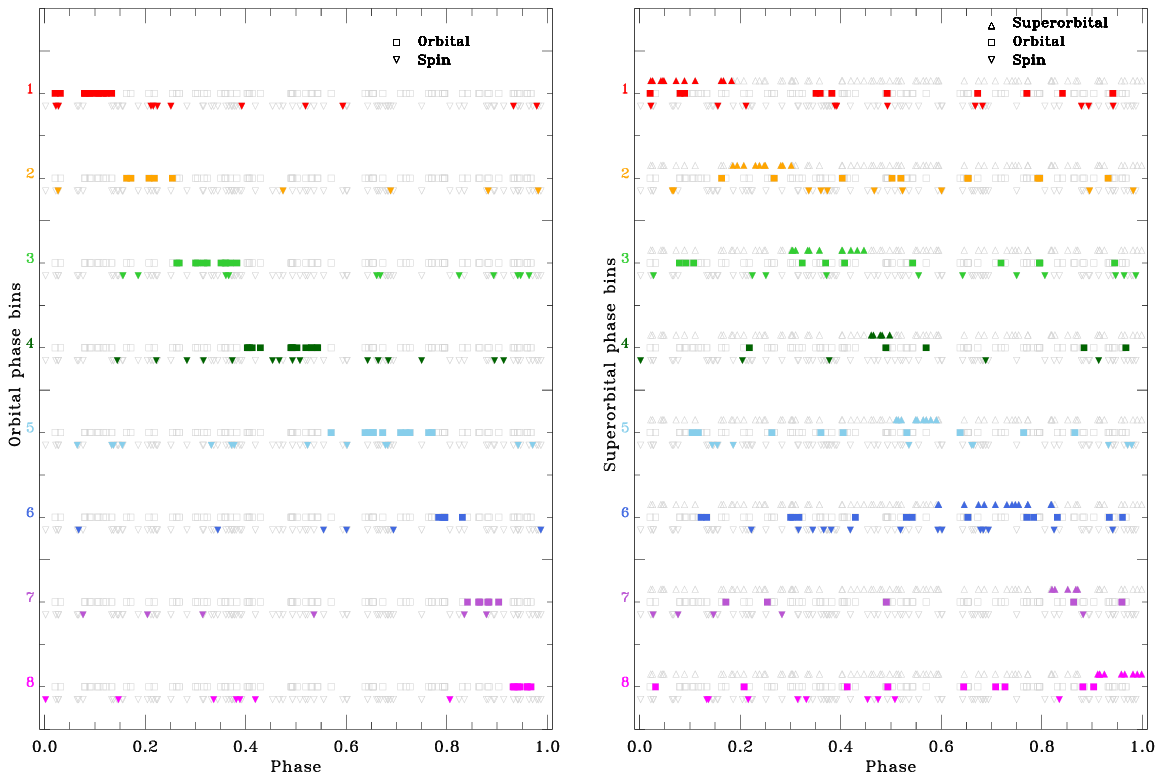}

  \vspace{-13.5truecm}
 \caption{{\it Left:} 
 Distribution of the observations collected on \mysrc\ (on the x axis) in 
 orbital phase (grey empty squares), 
 and spin phase (grey empty downward-pointing triangles) 
 as a function of orbital phase bin (on the y axis). 
 The observations for each orbital phase bin are shown in different colours as filled symbols and offset for clarity. 
  {\it Right:} 
 Distribution of the observations collected on \mysrc\ (on the x axis) in 
 superorbital phase (grey empty upward-pointing triangles), 
 orbital phase (grey empty squares), 
 and spin phase (grey empty downward-pointing triangles) 
 as a function of superorbital phase bin (on the y axis). 
 The observations for each superorbital phase bin are shown in different colours as filled symbols and offset for clarity.  
  \label{supero:fig:2S_3phases}}
 \end{figure*}

Figure~\ref{supero:fig:2S_3phases} (left) shows as grey empty squares where our sample of observations occur in the orbital phase 
and as grey downward-pointing empty triangles where they occur in spin phase (see, e.g.\, the first two rows). 
For each orbital phase bin 
(filled squares: red for the first bin, orange for the second bin, green for the third and so on) on the first row, 
we report as filled triangles the matching observations in spin phase in the second row (with the same colour). 
As can be observed, the spin phases, albeit being a small number, are not preferentially clustered. 
Therefore we can argue that, within the limits of our discrete sampling of the light curve of \mysrc, 
we are not introducing strong biases due to the distribution of our observations in spin phase 
in each orbital phase bin. 

Similarly, to address  the possibility that 
in each superorbital phase bin the orbital phases would not be equally represented, 
Figure~\ref{supero:fig:2S_3phases} (right) shows 
as grey upward-pointing empty triangles where our sample of observations occur in the superorbital phase, 
as grey empty squares where our sample of observations occur in the orbital phase 
and as grey downward-pointing empty triangles where they occur in spin phase (e.g., the first three rows). 
For each superorbital phase bin (filled triangles with the same colour scheme) on the first row, 
we report as filled symbols  the matching observations in orbital phase in the second row
and spin phase in the third row (with the same colour). 
Once more, we can argue that we are not introducing strong biases due to the distribution of our observations in orbital phase 
in each superorbital orbital phase bin.

We finally consider the other sources which data we reported in Table~\ref{supero:tab:sample}.  
In the case of \srca\ and \srcc\ the pulsation timescale is in the order of $\sim1$\,ks,  
comparable with the typical XRT exposure. The XRT data thus sample automatically all spin phases 
in each observation. 
For \srcb, no pulsations were ever detected and thus no such investigation is possible 
at this time. 

\vfill

%%%%%%%%%%%%%%%%%%%%%%%%%%%%%%%%%%%%%%%%%%%%%%%%%%%%%%%%%

% Don't change these lines
\bsp	% typesetting comment
\label{lastpage}
\end{document}